\documentclass[%reprint,
 amsmath,amssymb,aps,%fleqn
]{revtex4-1}

\usepackage{graphicx}% Include figure files
\usepackage{dcolumn}% Align table columns on decimal point
\usepackage{bm}% bold math
\usepackage{lipsum}

\newcommand{\mv}[1]{\ensuremath{\mathbf{#1}}} % for vectors
\newcommand{\gv}[1]{\ensuremath{\mbox{\boldmath$ #1 $}}} 
\newcommand{\abs}[1]{\left| #1 \right|} % for absolute value
\newcommand{\avg}[1]{\left \langle #1 \right \rangle} % for average
\newcommand{\der}[2]{\frac{d #1}{d #2}} % for derivatives
\newcommand{\pd}[2]{\frac{\partial #1}{\partial #2}} % for partial derivatives
\usepackage[colorlinks]{hyperref}  
\usepackage{cleveref}
\hypersetup{linkcolor=blue,citecolor=blue,urlcolor=blue}

\usepackage{titlesec}
\titleformat{\section}
  {\centering\normalfont\fontsize{11}{15}\bfseries}{\thesection}{1em}{}

\usepackage[toc]{appendix}

\crefname{figure}{fig.}{figs.}
\Crefname{figure}{Figures}{Figures}

\begin{document}

\setcounter{page}{1} %first page number

\title{Generalized Mode-Coupling Theory for Mixtures of Brownian Particles}

\author{Vincent E. Debets$^{1,2}$, Chengjie Luo$^{1,2}$, Simone Ciarella$^{1,3}$, Liesbeth M.C. Janssen$^{1,2*}$}

\affiliation{$^{1}$Department of Applied Physics, Eindhoven University of Technology, P.O. Box 513,
5600 MB Eindhoven, The Netherlands\\$^{2}$Institute for Complex
Molecular Systems, Eindhoven University of Technology, P.O. Box 513,
5600 MB Eindhoven, The Netherlands\\$^{3}$Laboratoire de Physique de l’Ecole Normale Supérieure, ENS, Université PSL, CNRS, Sorbonne Université, Université de Paris, F-75005 Paris, France\\}
\email{l.m.c.janssen@tue.nl}

\begin{abstract}%
{\noindent Generalized mode-coupling theory (GMCT) has recently emerged as a promising first-principles theory to study the poorly understood dynamics of glass-forming materials. Formulated as a hierarchical extension of standard mode-coupling theory (MCT), it is able to systematically improve its predictions by including the exact dynamics of higher-order correlation functions into its hierarchy.
However, in contrast to Newtonian dynamics, a fully generalized version of the theory based on Brownian dynamics is still lacking. To close this gap, we provide a detailed derivation of GMCT for colloidal mixtures obeying a many-body Smoluchowski equation. We demonstrate that a hierarchy of coupled equations can again be established and show that these, consistent with standard MCT, are \textit{identical} to the ones obtained from Newtonian GMCT when taking the overdamped limit. Consequently, the non-trivial similarity between Brownian and Newtonian MCT is maintained for our newly developed multi-component GMCT. As a proof of principle, we also solve the generalized mode-coupling equations for the binary Kob-Andersen Lennard-Jones mixture undergoing Brownian dynamics, and confirm the improved predictive power of the theory upon using more levels of the GMCT hierarchy of equations.}
\end{abstract}

\maketitle %%The above information typeset through this command

\section{Introduction}
\noindent To this date, the physical mechanisms underlying the liquid-to-glass transition are still not fully comprehended. 
%researchers struggle to fully comprehend  the nature of glass-forming materials and the process by which they are formed. 
Since glassy dynamics manifests itself in numerous systems across different length scales, e.g.\ atomic and molecular liquids, polymers, colloids, granular materials, and even living cells~\cite{Janssen2018front,Janssen2019active,Berthier2011theoretical,Klongvessa2019colloid1,Klongvessa2019colloid2,Mattsson2009,Angelini2011,Yang2021,Li2020,Abate2007,Ciarella2019pnas}, a better understanding of this process is pivotal to a wide range of research areas. The main problem, however, remains that when a material transitions from a liquid to a glassy or amorphous solid state, its viscosity or relaxation time shows a tremendous non-linear growth, while its structure and related thermodynamic control parameters exhibit only minor differences~\cite{debenedetti2001supercooled,Ediger1996}. Consequently, a successful theory for glassy dynamics not only needs to pick up on these subtle structural changes, but also has to adequately magnify them in order to accurately describe the dramatic dynamical slowdown. 

One of the few first-principles-based theories that is at least partially capable of this task is mode-coupling theory (MCT)~\cite{Janssen2018front,gotze2008complex,Gotze1992,reichman2005mode,Gotze1999}. This theory requires solely the static structure factor as input; from this structural information alone, it is able to predict the intermediate scattering function and thus the full microscopic relaxation dynamics of many glass-forming materials with qualitative and sometimes even semi-quantitative accuracy. Among its most notable successes are the prediction of two-step relaxation patterns, stretched exponentials, and universal scaling laws upon approaching the glass transition point~\cite{Janssen2018front,gotze2008complex,Gotze1992,reichman2005mode,Gotze1999,weysser2010structural}.

Unfortunately, MCT often fails to reach full quantitative accuracy; the primary reason for this resides in the theory's ad hoc Gaussian factorization and disregard of higher-order correlation functions. In recent years so-called generalized mode-coupling theory (GMCT) has therefore been introduced as a means to limit the effect of this uncontrolled approximation, with results that improve upon predictions of standard MCT in a systematic, seemingly convergent manner~\cite{JanssenPRE2014,Janssen2015a,Janssen2016a,Luo2020_1,Luo2020_2,ciarella2021,Mayer2006,SzamelPRL2003,Wu2005,Luo2021}. The basic notion of this theory is to develop separate equations of motion for higher-order correlation functions, instead of immediately applying a factorization approximation at the lowest possible order. This results in a hierarchy of MCT-like equations~\cite{JanssenPRE2014,Janssen2015a,Janssen2016a,Luo2020_1,Luo2020_2,ciarella2021,Luo2021}. So far, the explicit inclusion of more high-order correlators has been shown to significantly enhance predictions in comparison to MCT and near-quantitative accuracy in the weakly supercooled regime has already been reached~\cite{Janssen2015a}. Moreover, the theory has no free parameters, still only uses the static structure factor as input, and has thus presented itself as a promising first-principles theory to study the dynamics of glass-forming materials. 

However, most of the GMCT studies and the accompanying derivations have focused primarily on systems governed by Newtonian dynamics, while the few works based on Brownian dynamics are restricted to single-component systems and time-independent properties~\cite{SzamelPRL2003,Wu2005}. Considering that an appreciable amount of studied glassy materials, including polydisperse colloidal suspensions, is comprised of particles experiencing overdamped dynamics, it becomes apparent that a complete GMCT framework for multi-component systems governed by Brownian dynamics is desired. For standard MCT it is known that deriving the theory from either Newtonian or Brownian principles yields the same results~\cite{Nagele1999,ciarella2021}. This subtle and non-trivial equivalency might be attributed to the fact that in MCT all 'fast' variables are projected out and the differences between overdamped and underdamped dynamics seem to be covered within these 'fast' variables. Since GMCT retains this method of projecting out 'fast' variables and only extends it to higher order correlation functions, we expect the observed similarity to persist upon generalizing the theory, although this is not a priori evident. 

In this work, we seek to formally demonstrate this equivalency by deriving %, inspired by initial studies on colloidal systems already extending the framework of MCT~\cite{SzamelPRL2003,Wu2005}, 
a generalized MCT for colloidal mixtures obeying a many-body Smoluchowski equation. Our work significantly extends earlier GMCT studies~\cite{SzamelPRL2003,Wu2005} by accounting both for the full time-dependence of the Brownian relaxation dynamics, and for polydisperse or multiple-particle-species compositions. Below we first present our model system and briefly recap the results of standard MCT. Next, we provide detailed derivations of the equations of motion for the $4$-point and $6$-point density correlation functions and their dependence on the higher order $6$-point and $8$-point density correlation functions (via the memory kernels) respectively. We discuss the similarities between the newly developed equations of motion and the lowest order MCT equation of motion, and, based on them, extract a hierarchy of equations describing correlation functions of arbitrary order, thereby establishing a generalized MCT for mixtures of Brownian particles. A comparison is then made between Newtonian and Brownian GMCT, which are shown to be identical after taking the overdamped limit of the former. As a proof of principle, we also employ our theory to study some features, primarily the critical temperature, of the Kob-Andersen binary Lennard-Jones mixture~\cite{Kob1994}. This model system has been extensively studied in the context of glassy physics and earlier comparisons with MCT have shown that the theory strongly overestimates the critical temperature~\cite{Berthier2011,ciarella2021,Flenner2005,Nauroth1997}. We confirm the improved predictive power of GMCT by demonstrating that, upon including more levels of the hierarchy of equations, the critical temperature lowers in a seemingly convergent manner towards the simulation results.

\section{Theory}
\subsection*{Mode-Coupling Theory for Brownian Particles}
\noindent Our model glassy system is an $m$-component mixture consisting of $N$ interacting spherical Brownian particles immersed in a viscous medium of volume $V$. Following prior works on colloidal glasses, we neglect hydrodynamic interactions, so that the equation of motion for each particle is written as~\cite{Flenner2005}
\begin{equation}\label{eomBrownian}
    \der{\mv{r}^{\alpha}_{i}}{t} = \zeta_{\alpha}^{-1} \mv{F}^{\alpha}_{i}  + \gv{\xi}^{\alpha}_{i}.
\end{equation}
Here, $\mv{r}^{\alpha}_{i}$ denotes the position of the $i$th particle of type $\alpha$, $\mv{F}^{\alpha}_{i}$ the interaction force acting on it, $\zeta_{\alpha}$ the component-dependent friction constant, and $\gv{\xi}^{\alpha}_{i}$ a Gaussian thermal noise with zero mean and variance $\avg{\gv{\xi}^{\alpha}_{i}(t)\gv{\xi}^{\beta}_{j}(t^{\prime})}_{\mathrm{noise}}=2D^{\alpha}\mv{I}\delta_{ij}\delta_{\alpha \beta}\delta(t-t^{\prime})$, with $D^{\alpha}$ the thermal diffusion coefficient, $t$ the time, $\delta(t)$ the Dirac delta function, $\delta_{ij}$ the Kronecker delta, and $\mv{I}$ the unit matrix.
Based on these equations the joint $N$-particle probability density function (PDF) evolves in time via~\cite{Nagele1999,Nagele1996}
\begin{equation}
\pd{}{t}P_{N}(\mv{r}^{N},t)= \Omega P_{N}(\mv{r}^{N},t),
\end{equation}
where $\mv{r}^{N}=(\mv{r}_1^{1},...,\mv{r}_{N_m}^{m})$ denotes the configuration space and the evolution (or Smoluchowski) operator $\Omega$ is given by
\begin{equation}
    \Omega = \sum_{\alpha=1}^{m}\sum_{i=1}^{N_{\alpha}}  D^{\alpha} \nabla^{\alpha}_{i} \cdot \left( \nabla^{\alpha}_{i} - \beta  \mv{F}^{\alpha}_{i}  \right),
\end{equation}
with $N_{\alpha}$ the number of particles of type $\alpha$ and $\beta$ the inverse thermal energy. In equilibrium the PDF yields the Boltzmann solution 
$P_{\mathrm{eq}}(\mv{r}^{N})\propto \exp\left(-\beta U(\mv{r}^{N})\right)$. This solution depends solely on the total interaction potential $U(\mv{r}^{N})$, which is spherically symmetric and produces the interaction forces $\mv{F}^{\alpha}_{i}=-\nabla^{\alpha}_{i}U(\mv{r}^{N})$. Moreover, the friction constant and inverse thermal energy are related to each other via the Stokes-Einstein equation: $\beta D^{\alpha} = \zeta_{\alpha}^{-1}$. Note that $D^{\alpha}$ represents the intrinsic short-time diffusivity of the particles and should not be confused with the long-time diffusion coefficient, which may in fact violate the Stokes-Einstein equation upon approaching the glass transition point~\cite{weysser2010structural,Flenner2005,tarjus1995,shi2013,kumar2006}. 

The collective motion of particles can then be described via density modes: 
\begin{equation}
    \rho_{\alpha} (\mv{k})=\frac{1}{\sqrt{N_{\alpha}}}\sum_{j=1}^{N_{\alpha}} e^{i\mv{k}\cdot \mv{r}^{\alpha}_{j}},
\end{equation}
with wavevector $\mv{k}$ and the factor $1/\sqrt{N_{\alpha}}$ added for normalization. 
Our main probes to study the cooperative diffusion and glassy behavior of the colloids are the so-called partial dynamic structure factors (or intermediate scattering functions), which are the time-correlation between two such density modes:
\begin{equation}
    F_{\alpha;\beta}(k,t)=\avg{\rho_{\alpha}^{*} (\mv{k}) e^{\Omega t} \rho_{\beta} (\mv{k})}.
    \label{eq:F}
\end{equation}
Note that they only depend on the absolute value $k=\abs{\mv{k}}$ due to the system being isotropic, and that, as a result of translational invariance, they are non-zero only when the wavevectors in both density modes are equal. At time zero the time-correlation in \cref{eq:F} defines the static structure factor:
\begin{equation}
    F_{\alpha;\beta}(k,t=0)\equiv S_{\alpha;\beta}(k) =  \avg{\rho_{\alpha}^{*} (\mv{k}) \rho_{\beta} (\mv{k})}.
\end{equation}
We point out that averaging $\avg{...}$ is done with respect to the equilibrium distribution $P_{\mathrm{eq}}(\mv{r}^{N})$ and that the operator $\Omega$ works on everything to its right including the PDF.

Applying the Mori-Zwanzig projector operator formalism~\cite{Mori65,Zwanzig60} the time evolution of $F_{\alpha;\beta}(k,t)$ can be shown to obey the following equation~\cite{Nagele1999,Flenner2005}
\begin{equation}
    \pd{}{t}F_{\alpha;\beta}(k,t) + H_{\alpha;\gamma}(k) S^{-1}_{\gamma;\epsilon}(k) F_{\epsilon;\beta}(k,t) + \int_{0}^{t} dt^{\prime}  M_{\alpha;\gamma}(k,t-t^{\prime}) H^{-1}_{\gamma;\epsilon}(k) \pd{}{t^{\prime}}F_{\epsilon;\beta}(k,t^{\prime}) =0,
    \label{eq:bMCT}
\end{equation}
where we have adopted the Einstein summation convention to sum over repeated indices, and the 
superscript $-1$ denotes the inverse matrix of the respective quantity, i.e.\ $X^{-1}_{\gamma;\epsilon}\equiv [\mathbf{X}^{-1}]_{\gamma;\epsilon}$.
The collective $2$-point diffusion matrix governing the short-time dynamics is given by
\begin{equation}
    H_{\alpha;\beta}(k) = -\avg{\rho_{\alpha}^{*}  (\mv{k})\Omega \rho_{\beta} (\mv{k})}= k^{2} D^{\alpha}\delta_{\alpha\beta}.
\end{equation}
The crucial term of \cref{eq:bMCT} is the irreducible $2$-point memory kernel $M_{\alpha;\beta}(k,t)$, which contains the time-autocorrelation function of the fluctuating forces. In MCT, these fluctuating forces are projected onto the subspace of density doublets, which are assumed to be the slowest modes after the single density modes~\cite{reichman2005mode}). By subsequently invoking the convolution approximation~\cite{Jackson1962} and Gaussian factorization~\cite{Janssen2015a} for the 3-point and 4-point static density correlations, respectively, which should be reasonable for systems not prone to form networks~\cite{Sciortino2001}, the memory kernel can be written as~\cite{Nagele1999,Szamel1991}
\begin{equation}\label{M1exact}
    M_{\alpha;\beta}(k,t)=\frac{1}{4}\sum_{\mv{q},\mv{q}^{\prime}} \frac{D^{\alpha}}{\sqrt{N_{\alpha}}} V^{\alpha}_{\mu\nu}(\mv{k},\mv{q}) \avg{\rho_{\mu}^{*} (\mv{q})\rho_{\nu}^{*} (\mv{k}-\mv{q}) e^{\Omega_{\mathrm{irr}} t} \rho_{\mu^{\prime}} (\mv{q}^{\prime})\rho_{\nu^{\prime}} (\mv{k}-\mv{q}^{\prime})} \frac{D^{\beta}}{\sqrt{N_{\beta}}} V^{\beta}_{\mu^{\prime}\nu^{\prime}}(\mv{k},\mv{q}^{\prime}),
\end{equation}
with $\Omega_{\mathrm{irr}}$ representing the irreducible evolution operator~\cite{Nagele1999,Cichocki,Szamel1991,Kawasaki1994}. The vertices, which represent the coupling strength between different wavenumbers, are written in terms of the direct correlation function $C_{\alpha\beta}(k)=\delta_{\alpha\beta}-S_{\alpha\beta}^{-1}(k)$ as
\begin{equation}
    V^{\alpha}_{\mu\nu}(\mv{k},\mv{q})= \mv{k} \cdot \mv{q}\ \delta_{\alpha \nu} C_{\alpha \mu}(q) + \mv{k} \cdot (\mv{k}-\mv{q})\ \delta_{\alpha \mu} C_{\alpha \nu}(\abs{\mv{k}-\mv{q}}).
\end{equation}
It is apparent that the irreducible $2$-point memory kernel requires an even more complex (unknown) irreducible $4$-point dynamic density correlation, which prevents one from finding numerical solutions. Approximations are therefore desired and in standard MCT this takes the shape of an ad hoc factorization. In particular, the irreducible evolution operator $\Omega_{\mathrm{irr}}$ is replaced by the full time evolution operator $\Omega$ and the $4$-point density correlations are written as products of $2$-point density correlations, yielding~\cite{Nagele1999,Szamel1991}  
\begin{equation}\label{MCT_approx}
    \avg{\rho_{\mu}^{*} (\mv{q})\rho_{\nu}^{*} (\mv{k}-\mv{q}) e^{\Omega^{\mathrm{irr}} t} \rho_{\mu^{\prime}} (\mv{q}^{\prime})\rho_{\nu^{\prime}} (\mv{k}-\mv{q}^{\prime})}\approx F_{\mu;\mu^{\prime}}(q,t)F_{\nu;\nu^{\prime}}(\abs{\mv{k}-\mv{q}},t)\delta_{\mv{q},\mv{q}^{\prime}} + F_{\mu;\nu^{\prime}}(q,t)F_{\nu;\mu^{\prime}}(\abs{\mv{k}-\mv{q}},t)\delta_{\mv{k}-\mv{q},\mv{q}^{\prime}}.
\end{equation}
Consequently, the irreducible $2$-point memory kernel simplifies to~\cite{Nagele1999}
\begin{equation}
    M_{\alpha;\beta}(k,t)=\frac{1}{2}\sum_{\mv{q}} \frac{D^{\alpha}}{\sqrt{N_{\alpha}}} V^{\alpha}_{\mu\nu}(\mv{k},\mv{q}) F_{\mu;\mu^{\prime}}(q,t)F_{\nu;\nu^{\prime}} (\abs{\mv{k}-\mv{q}},t) \frac{D^{\beta}}{\sqrt{N_{\beta}}} V^{\beta}_{\mu^{\prime}\nu^{\prime}}(\mv{k},\mv{q}),
\end{equation}
which, using the static structure factor $F_{\alpha;\beta}(k,t=0)=S_{\alpha;\beta}(k)$ as the initial boundary condition, allows us to self-consistently find a solution for $F_{\alpha;\beta}(k,t)$ and study the glassy dynamics of our colloidal mixture.

\subsection*{Extending towards GMCT}
\noindent Despite remarkable successes of MCT~\cite{Kob2002,weysser2010structural,reichman2005mode,Janssen2018front}, several discrepancies between the theory and simulations or experiments still persist~\cite{Berthier2010,Berthier2011,Banerjee2014,Dell2015,Landes2020}, most notably an overestimation of the glass transition. Inspired by previous studies~\cite{SzamelPRL2003,Wu2005,Mayer2006, JanssenPRE2014,Janssen2015a,Janssen2016a,Luo2020_1,Luo2020_2,ciarella2021}, we therefore seek to improve and generalize the theory by developing a separate equation for the $4$-point dynamic density correlations (and later on continuing the process for even higher order correlations). Since we believe the factorization to be the most severe approximation, we will retain the $4$-point density correlations and only replace the irreducible evolution operator by a full one. Moreover, to keep calculations tractable and consistent with previous work on Newtonian GMCT, we will also apply the diagonal approximation ($\mv{q}=\mv{q}^{\prime}$,\  $\mv{k}-\mv{q}=\mv{q}^{\prime}$)~\cite{Janssen2015a,Luo2020_1,Luo2020_2}, so that overall we have
\begin{equation}
    \avg{\rho_{\mu}^{*} (\mv{q})\rho_{\nu}^{*} (\mv{k}-\mv{q}) e^{\Omega^{\mathrm{irr}} t} \rho_{\mu^{\prime}} (\mv{q}^{\prime})\rho_{\nu^{\prime}} (\mv{k}-\mv{q}^{\prime})}\approx F^{(2)}_{\mu\nu;\mu^{\prime}\nu^{\prime}}(q,\abs{\mv{k}-\mv{q}},t)\delta_{\mv{q},\mv{q}^{\prime}}+F^{(2)}_{\mu\nu;\nu^{\prime}\mu^{\prime}}(q,\abs{\mv{k}-\mv{q}},t) \delta_{\mv{k}-\mv{q},\mv{q}^{\prime}}.
\end{equation}
As a result, the memory kernel becomes a function of the diagonal $4$-point dynamic density correlation function $\\F^{(2)}_{\alpha_{1}\alpha_{2};\beta_{1}\beta_{2}}(k_{1},k_{2},t) =\avg{\rho_{\alpha_{1}}^{*} (\mv{k}_{1})\rho_{\alpha_{2}}^{*} (\mv{k}_{2}) e^{\Omega t} \rho_{\beta_{1}} (\mv{k}_{1})\rho_{\beta_{2}} (\mv{k}_{2})}$:
\begin{equation}\label{M1final}
    M_{\alpha;\beta}(k,t)=\frac{1}{2}\sum_{\mv{q}} \frac{D^{\alpha}}{\sqrt{N_{\alpha}}} V^{\alpha}_{\mu\nu}(\mv{k},\mv{q}) F^{(2)}_{\mu\nu;\mu^{\prime}\nu^{\prime}}(q,\abs{\mv{k}-\mv{q}},t) \frac{D^{\beta}}{\sqrt{N_{\beta}}} V^{\beta}_{\mu^{\prime}\nu^{\prime}}(\mv{k},\mv{q}).
\end{equation}

Having simplified the irreducible to a diagonal $4$-point density correlation, we can now once more resort to the Mori-Zwanzig projection operator formalism to describe its dynamics. Due to the diagonal form of $F^{(2)}_{\alpha_{1}\alpha_{2};\beta_{1}\beta_{2}}(k_{1},k_{2},t)$, we will employ the following  projection operator on the subspace defined by density doublets,
\begin{equation}
\mathcal{P}^{(2)}= |  \rho_{\alpha_{1}} (\mv{k}_{1})\rho_{\alpha_{2}} (\mv{k}_{2}) \rangle \chi^{(2)}_{12}
\langle \rho_{\beta_{1}}^{*} (\mv{k}_{1})\rho_{\beta_{2}}^{*} (\mv{k}_{2}) | ,
\end{equation}
and its orthogonal complement $\mathcal{Q}^{(2)}=\mathcal{I}-\mathcal{P}^{(2)}$, with the normalization factor $\chi^{(2)}_{12}$ ensuring the idempotency of the projection, i.e.\ $\mathcal{P}^{(2)}\mathcal{P}^{(2)}=\mathcal{P}^{(2)}$. One can then derive an exact equation of motion for the $4$-point density correlation functions (see supplementary information for more details), which is given by
\begin{equation}\label{eom_F2_K}
    \begin{split}
    \pd{}{t}F^{(2)}_{\alpha_{1}\alpha_{2};\beta_{1}\beta_{2}}(k_{1},k_{2},t) & + H^{(2)}_{\alpha_{1}\alpha_{2};\gamma_{1}\gamma_{2}}(k_{1},k_{2}) \left(S^{(2)}\right)^{-1}_{\gamma_{1}\gamma_{2};\epsilon_{1}\epsilon_{2}}(k_{1},k_{2})\  F^{(2)}_{\epsilon_{1}\epsilon_{2};\beta_{1}\beta_{2}}(k_{1},k_{2},t)\\
    &- \int_{0}^{t} dt^{\prime}  K^{(2)}_{\alpha_{1}\alpha_{2};\gamma_{1}\gamma_{2}}(k_{1},k_{2},t-t^{\prime}) \left(S^{(2)}\right)^{-1}_{\gamma_{1}\gamma_{2};\epsilon_{1}\epsilon_{2}}(k_{1},k_{2})\  \pd{}{t^{\prime}}F^{(2)}_{\epsilon_{1}\epsilon_{2};\beta_{1}\beta_{2}}(k_{1},k_{2},t^{\prime}) =0.
    \end{split}
\end{equation}
In this equation the collective diffusion matrix and static structure factor are naturally extended towards their 4-point tensorial counterparts:
\begin{equation}
    H^{(2)}_{\alpha_{1}\alpha_{2};\beta_{1}\beta_{2}}(k_{1},k_{2})= - \avg{\rho_{\alpha_{1}}^{*} (\mv{k}_{1})\rho_{\alpha_{2}}^{*} (\mv{k}_{2})\Omega \rho_{\beta_{1}} (\mv{k}_{1})\rho_{\beta_{2}} (\mv{k}_{2})}=k_{1}^{2} D^{\alpha_{1}}\delta_{\alpha_{1}\beta_{1}}S_{\alpha_{2};\beta_{2}}(k_{2})+k_{2}^{2}D^{\alpha_{2}}\delta_{\alpha_{2}\beta_{2}}S_{\alpha_{1};\beta_{1}}(k_{1}),
\end{equation}
and
\begin{equation*}
   S^{(2)}_{\alpha_{1}\alpha_{2};\beta_{1}\beta_{2}}(k_{1},k_{2})=\avg{\rho_{\alpha_{1}}^{*} (\mv{k}_{1})\rho_{\alpha_{2}}^{*} (\mv{k}_{2}) \rho_{\beta_{1}} (\mv{k}_{1})\rho_{\beta_{2}} (\mv{k}_{2})}\approx S_{\alpha_{1};\beta_{1}}(k_{1}) S_{\alpha_{2};\beta_{2}}(k_{2}),
\end{equation*}
where the latter is approximated using Gaussian factorization, which, from this point onward, will be done throughout for static correlations. The inverse operation represented by the superscript $-1$ is defined via
\begin{equation}\label{inv_def}
    S^{(2)}_{\alpha_{1}\alpha_{2};\gamma_{1}\gamma_{2}}(k_{1},k_{2}) \left(S^{(2)}\right)^{-1}_{\gamma_{1}\gamma_{2};\beta_{1}\beta_{2}}(k_{1},k_{2}) = \delta_{\alpha_{1}\beta_{1}}\delta_{\alpha_{2}\beta_{2}}.
\end{equation}
This implies that the inverse 4-point static structure tensor, i.e.\ the normalization factor $\chi^{(2)}_{12}$, can be written as
\begin{equation}
    \chi^{(2)}_{12}=\left(S^{(2)}\right)^{-1}_{\alpha_{1}\alpha_{2};\beta_{1}\beta_{2}}(k_{1},k_{2}) \approx S^{-1}_{\alpha_{1};\beta_{1}}(k_{1}) S^{-1}_{\alpha_{2};\beta_{2}}(k_{2}).
\end{equation}
Under the assumption of static Gaussian factorization we thus see that inverse higher order structure factors are equally factorized into inverse $2$-point structure factors.
The $4$-point memory kernel captures all nontrivial dynamics and is formally written as
\begin{equation}
    K^{(2)}_{\alpha_{1}\alpha_{2};\beta_{1}\beta_{2}}(k_{1},k_{2},z)=\avg{\rho_{\alpha_{1}}^{*} (\mv{k}_{1})\rho_{\alpha_{2}}^{*} (\mv{k}_{2}) \Omega \mathcal{Q}^{(2)} (z-\mathcal{Q}^{(2)}\Omega \mathcal{Q}^{(2)})^{-1} \mathcal{Q}^{(2)} \Omega \rho_{\beta_{1}} (\mv{k}_{1})\rho_{\beta_{2}} (\mv{k}_{2})}.
\end{equation}
However, in its present form the memory kernel does not lend itself to
MCT-like approximations~\cite{Cichocki,Kawasaki1994} and we therefore follow standard procedure by converting it to an irreducible memory kernel. Employing a further projection operator,
\begin{equation}
\mathcal{P}_{\mathrm{irr}}^{(2)}=  - | \rho_{\alpha_{1}} (\mv{k}_{1})\rho_{\alpha_{2}} (\mv{k}_{2}) \rangle \left(H^{(2)}\right)^{-1}_{\alpha_{1}\alpha_{2};\beta_{1}\beta_{2}}(k_{1},k_{2})\ 
\langle \rho_{\beta_{1}}^{*} (\mv{k}_{1})\rho_{\beta_{2}}^{*} (\mv{k}_{2}) \Omega | , \hspace{1.0cm} \mathcal{Q}_{\mathrm{irr}}^{(2)}= \mathcal{I} - \mathcal{P}_{\mathrm{irr}}^{(2)},
\end{equation}
we find that (see supplementary information)
\begin{equation}
    K^{(2)}_{\alpha_{1}\alpha_{2};\beta_{1}\beta_{2}}(k_{1},k_{2},t)=M^{(2)}_{\alpha_{1}\alpha_{2};\beta_{1}\beta_{2}}(k_{1},k_{2},t) - \int_{0}^{t}dt^{\prime}M^{(2)}_{\alpha_{1}\alpha_{2};\gamma_{1}\gamma_{2}}(k_{1},k_{2},t-t^{\prime})\left(H^{(2)}\right)^{-1}_{\gamma_{1}\gamma_{2};\epsilon_{1}\epsilon_{2}}(k_{1},k_{2})\ K^{(2)}_{\epsilon_{1}\epsilon_{2};\beta_{1}\beta_{2}}(k_{1},k_{2},t^{\prime}).
\end{equation}
This can then be combined with \cref{eom_F2_K} to arrive at an almost equivalent equation of motion for the 4-point density correlation function as the one introduced earlier for the 2-point density correlation function:
\begin{equation}\label{eom_F2}
    \begin{split}
    \pd{}{t}F^{(2)}_{\alpha_{1}\alpha_{2};\beta_{1}\beta_{2}}(k_{1},k_{2},t) & + H^{(2)}_{\alpha_{1}\alpha_{2};\gamma_{1}\gamma_{2}}(k_{1},k_{2}) \left(S^{(2)}\right)^{-1}_{\gamma_{1}\gamma_{2};\epsilon_{1}\epsilon_{2}}(k_{1},k_{2})\  F^{(2)}_{\epsilon_{1}\epsilon_{2};\beta_{1}\beta_{2}}(k_{1},k_{2},t)\\
    &+ \int_{0}^{t} dt^{\prime}  M^{(2)}_{\alpha_{1}\alpha_{2};\gamma_{1}\gamma_{2}}(k_{1},k_{2},t-t^{\prime}) \left(H^{(2)}\right)^{-1}_{\gamma_{1}\gamma_{2};\epsilon_{1}\epsilon_{2}}(k_{1},k_{2})\  \pd{}{t^{\prime}}F^{(2)}_{\epsilon_{1}\epsilon_{2};\beta_{1}\beta_{2}}(k_{1},k_{2},t^{\prime}) =0.
    \end{split}
\end{equation}
Here, the 4-point irreducible memory kernel reads 
\begin{equation}\label{M2_formal}
    M^{(2)}_{\alpha_{1}\alpha_{2};\beta_{1}\beta_{2}}(k_{1},k_{2},t)=\avg{\rho_{\alpha_{1}}^{*} (\mv{k}_{1})\rho_{\alpha_{2}}^{*} (\mv{k}_{2})\Omega \mathcal{Q}^{(2)}e^{\Omega^{(2)}_{\mathrm{irr}}t}\mathcal{Q}^{(2)}\Omega \rho_{\beta_{1}} (\mv{k}_{1})\rho_{\beta_{2}} (\mv{k}_{2})},
\end{equation}
and the corresponding irreducible evolution operator with respect to density doublets is defined as $\Omega^{(2)}_{\mathrm{irr}}=\mathcal{Q}^{(2)}\Omega\mathcal{Q}_{\mathrm{irr}}^{(2)}\mathcal{Q}^{(2)}$. Unfortunately, the complexity of this operator hinders any analytical progress and our strategy to develop Brownian-GMCT is therefore to simplify it by invoking a similar strategy as is done in MCT. Realizing that the second order fluctuating forces consist, to leading order, of products of three density modes~\cite{Janssen2015a,SzamelPRL2003}, we seek to project the second order fluctuating forces onto the subset of density triplets. More concretely, this translates to replacing $\langle \rho_{\alpha_{1}}^{*} (\mv{k}_{1})\rho_{\alpha_{2}}^{*} (\mv{k}_{2})\Omega \mathcal{Q}^{(2)} \rightarrow \langle \rho_{\alpha_{1}}^{*} (\mv{k}_{1})\rho_{\alpha_{2}}^{*} (\mv{k}_{2})\Omega \mathcal{Q}^{(2)}\mathcal{P}_{3}$ and $ \mathcal{Q}^{(2)}\Omega\rho_{\beta_{1}} (\mv{k}_{1})\rho_{\beta_{2}} (\mv{k}_{2}) \rangle \rightarrow \mathcal{P}_{3} \mathcal{Q}^{(2)}\Omega\rho_{\beta_{1}} (\mv{k}_{1})\rho_{\beta_{2}} (\mv{k}_{2}) \rangle$ in the $4$-point irreducible memory kernel, using the projection operator
\begin{equation}
    \mathcal{P}_{3}=\frac{1}{6}\sum_{\mv{q}_{1},\mv{q}_{2},\mv{q}_{3}} | \rho_{\mu_{1}} (\mv{q}_{1})\rho_{\mu_{2}} (\mv{q}_{2}) \rho_{\mu_{3}} (\mv{q}_{3}) \rangle \chi_{123}
\langle \rho_{\nu_{1}}^{*} (\mv{q}_{1})\rho_{\nu_{2}}^{*} (\mv{q}_{2}) \rho_{\nu_{3}}^{*} (\mv{q}_{3}) |.
\end{equation}
Note that the factor $1/6$ corrects for double counting and that the normalization $\chi_{123}=S^{-1}_{\mu_{1};\nu_{1}}(q_{1})  S^{-1}_{\mu_{2};\nu_{2}}(q_{2})
S^{-1}_{\mu_{3};\nu_{3}}(q_{3})$ is chosen in accordance with Gaussian factorization. Due to this projection, the memory kernel takes on the more familiar form of two 'vertices' enclosing a correlation one step above in the hierarchy, which in this case is a $6$-point irreducible correlation function 
\begin{equation}
\begin{split}
    & M^{(2)}_{\alpha_{1}\alpha_{2};\beta_{1}\beta_{2}}  (k_{1},k_{2},t)  \approx \frac{1}{36} \sum_{\mv{q}_{1}...\mv{q}_{6}} \avg{\rho_{\alpha_{1}}^{*} (\mv{k}_{1})\rho_{\alpha_{2}}^{*} (\mv{k}_{2})\Omega \mathcal{Q}^{(2)} \rho_{\mu_{1}} (\mv{q}_{1})\rho_{\mu_{2}} (\mv{q}_{2}) \rho_{\mu_{3}}(\mv{q}_{3})}\chi_{123} \\ & \avg{\rho_{\nu_{1}}^{*} (\mv{q}_{1})\rho_{\nu_{2}}^{*} (\mv{q}_{2}) \rho_{\nu_{3}}^{*} (\mv{q}_{3})e^{\Omega^{(2)}_{\mathrm{irr}}t} \rho_{\mu_{4}} (\mv{q}_{4})\rho_{\mu_{5}} (\mv{q}_{5}) \rho_{\mu_{6}} (\mv{q}_{6})} \chi_{456} \avg{\rho^{*}_{\nu_{4}} (\mv{q}_{4})\rho^{*}_{\nu_{5}} (\mv{q}_{5}) \rho^{*}_{\nu_{6}}(\mv{q}_{6})\Omega \mathcal{Q}^{(2)} \rho_{\beta_{1}} (\mv{k}_{1})\rho_{\beta_{2}} (\mv{k}_{2})}.
\end{split}
\end{equation}
Following the detailed derivation reported in the supplementary information, it is possible to reduce the left vertex to
\begin{equation} \label{vertex2}
\begin{split}
    \avg{\rho_{\alpha_{1}}^{*} (\mv{k}_{1})\rho_{\alpha_{2}}^{*} (\mv{k}_{2}) \Omega \mathcal{Q}^{(2)} \rho_{\mu_{1}} (\mv{q}_{1})\rho_{\mu_{2}} (\mv{q}_{2}) \rho_{\mu_{3}}(\mv{q}_{3})} \chi_{123} = & \Bigg[ \bigg(\frac{D^{\alpha_{1}}}{\sqrt{N_{\alpha_{1}}}}  \delta_{\mv{k}_{1},\mv{q}_{1}+\mv{q}_{2}}\delta_{\mv{k}_{2},\mv{q}_{3}} \delta_{\alpha_{2}\nu_{3}}  \Big(  k^{2}_{1} \delta_{\alpha_{1}\nu_{1}} \delta_{\alpha_{1}\nu_{2}} +  \mv{k}_{1}\cdot\mv{q}_{1} \delta_{\alpha_{1}\nu_{2}}
    S^{-1}_{\alpha_{1};\nu_{1}}(q_{1}) \\
    & \hspace{-5.0cm} +\ \mv{k}_{1}\cdot\mv{q}_{2} \delta_{\alpha_{1}\nu_{1}} S^{-1}_{\alpha_{1};\nu_{2}}(q_{2}) \Big) \bigg) \ +\ \bigg( \{ \mv{q}_{1},\nu_{1} \leftrightarrow \mv{q}_{3},\nu_{3}\} \bigg)  +  \bigg( \{ \mv{q}_{2},\nu_{2} \leftrightarrow \mv{q}_{3},\nu_{3}\} \bigg) \Bigg] + \Bigg[ \{ \mv{k}_{1},\alpha_{1} \leftrightarrow \mv{k}_{2},\alpha_{2}\} \Bigg],
\end{split}
\end{equation}
while the right vertex yields an identical expression. Here, the double-arrow contributions correspond to the aforementioned terms enclosed by the same brackets except for a swapping of the indicated wavevectors and particle labels. Inserting both results, evaluating the Kronecker deltas, and exploiting the symmetry of the different vertex terms allows us to rewrite the memory kernel as
\begin{equation}
\begin{split}
    &M^{(2)}_{\alpha_{1}\alpha_{2};\beta_{1}\beta_{2}}  (k_{1},k_{2},t) \approx  \frac{1}{4} \sum_{\mv{q},\mv{q}^{\prime}} \Bigg[ \frac{D^{\alpha_{1}}}{\sqrt{N_{\alpha_{1}}}} V^{\alpha_{1}}_{\mu\nu}(\mv{k}_{1},\mv{q})  \bigg( \avg{\rho_{\mu}^{*} (\mv{q})\rho_{\nu}^{*} (\mv{k}_{1}-\mv{q})\rho_{\alpha_{2}}^{*} (\mv{k}_{2}) e^{\Omega^{(2)}_{\mathrm{irr}} t} \rho_{\mu^{\prime}} (\mv{q}^{\prime})\rho_{\nu^{\prime}} (\mv{k}_{1}-\mv{q}^{\prime})\rho_{\beta_{2}} (\mv{k}_{2})} \frac{D^{\beta_{1}}}{\sqrt{N_{\beta_{1}}}} \\
    & V^{\beta_{1}}_{\mu^{\prime}\nu^{\prime}}(\mv{k}_{1},\mv{q}^{\prime}) + \avg{\rho_{\mu}^{*} (\mv{q})\rho_{\nu}^{*} (\mv{k}_{1}-\mv{q})\rho_{\alpha_{2}}^{*} (\mv{k}_{2}) e^{\Omega^{(2)}_{\mathrm{irr}} t} \rho_{\mu^{\prime}} (\mv{q}^{\prime})\rho_{\nu^{\prime}} (\mv{k}_{2}-\mv{q}^{\prime})\rho_{\beta_{1}} (\mv{k}_{1})} \frac{D^{\beta_{2}}}{\sqrt{N_{\beta_{2}}}} V^{\beta_{2}}_{\mu^{\prime}\nu^{\prime}}(\mv{k}_{2},\mv{q}^{\prime}) \bigg) \Bigg] + \Bigg[ \{ \mv{k}_{1},\alpha_{1},\beta_{1} \leftrightarrow \mv{k}_{2},\alpha_{2},\beta_{2}\} \Bigg],
\end{split}
\end{equation}
which is very reminiscent of \cref{M1exact} and therefore lends itself to similar approximations. In particular, we can again replace the irreducible operator with a full evolution operator and apply the diagonal approximation, which for second order comes down to $\mv{q}=\mv{q}^{\prime},\ \mv{k}_{1,2}-\mv{q}=\mv{q}^{\prime}$, and retaining only diagonal correlations~\cite{Janssen2015a,Luo2020_1,Luo2020_2}. Combined with the assumption that $\mv{k}_{1}\neq \mv{k}_{2}$, the memory kernel finally reduces to 
\begin{equation}\label{M2final}
    M^{(2)}_{\alpha_{1}\alpha_{2};\beta_{1}\beta_{2}}(k_{1},k_{2},t)\approx\frac{1}{2}\sum_{\mv{q}} \bigg( \frac{D^{\alpha_{1}}}{\sqrt{N_{\alpha_{1}}}} V^{\alpha_{1}}_{\mu\nu}(\mv{k}_{1},\mv{q}) F^{(3)}_{\mu\nu\alpha_{2};\mu^{\prime}\nu^{\prime}\beta_{2}}(q,\abs{\mv{k}_{1}-\mv{q}},k_{2},t) \frac{D^{\beta_{1}}}{\sqrt{N_{\beta_{1}}}} V^{\beta_{1}}_{\mu^{\prime}\nu^{\prime}}(\mv{k}_{1},\mv{q}) \bigg) \
    +\ \bigg( \{ \mv{k}_{1},\alpha_{1},\beta_{1} \leftrightarrow \mv{k}_{2},\alpha_{2},\beta_{2}\} \bigg),
\end{equation}
and thus becomes a function of the diagonal $6$-point density correlation
\begin{equation}
F^{(3)}_{\alpha_{1}\alpha_{2}\alpha_{3};\beta_{1}\beta_{2}\beta_{3}}(k_{1},k_{2},k_{3},t) =\avg{\rho_{\alpha_{1}}^{*} (\mv{k}_{1})\rho_{\alpha_{2}}^{*} (\mv{k}_{2}) \rho_{\alpha_{3}}^{*} (\mv{k}_{3}) e^{\Omega t} \rho_{\beta_{1}} (\mv{k}_{1})\rho_{\beta_{2}} (\mv{k}_{2})\rho_{\beta_{3}} (\mv{k}_{3})}.
\end{equation}
To recapitulate, we now have a set of two coupled equations for the $2$- and $4$-point density correlations, ~\cref{eq:bMCT,eom_F2}, which can be solved self-consistently by using an approximate expression for the $6$-point density correlation function, using as input the static structure factor and the initial boundary conditions. Normally, this is done by factorizing the $6$-point correlation function in terms of its lower order counterparts \cite{SzamelPRL2003,Janssen2015a}. This system of two coupled equations constitutes a first improvement upon standard MCT, since it pushes the factorization approximation back to a higher order correlation function, hence the name second order GMCT.

\subsection*{Hierarchy of Equations}
\noindent We are, however, by no means forced to introduce a factorization at the level of the $6$-point density correlations. In fact, the entire process laid out for the $4$-point density correlation function can be equally repeated for the diagonal $6$-point density correlations. This yields an almost identical equation of motion for $F^{(3)}_{\alpha_{1}\alpha_{2}\alpha_{3};\beta_{1}\beta_{2}\beta_{3}}(k_{1},k_{2},k_{3},t)$ (see \cref{eomF3}), which, after simplifying the irreducible 6-point memory kernel, can be shown to depend on the diagonal 8-point density correlation functions (see supplementary information for more details). An inspection of the simplified second [\cref{M2final}] and third [\cref{M3final}] order irreducible memory kernels then shows that they are very similar and we may, in agreement with previous work on GMCT, observe a hierarchy of equations starting to unfold itself~\cite{JanssenPRE2014,Janssen2015a,Janssen2016a,Luo2020_1,Luo2020_2,ciarella2021,Luo2021,Biezemans2020}. Extrapolating the observed behavior, we define a general $2n$-density correlation function
\begin{equation}
    F^{(n)}_{\{\alpha_{i}\};\{\beta_{i}\}}(\{k_{i}\},t) = \left\langle \rho^{*}_{\alpha_{1}}(\mv{k}_{1})\hdots \rho^{*}_{\alpha_{n}}(\mv{k}_{n}) e^{\Omega t}\rho_{\beta_{1}}(\mv{k}_{1})\hdots \rho_{\beta_{n}}(\mv{k}_{n}) \right\rangle,
\end{equation}
which obeys the following equation of motion
\begin{equation}\label{Fn_eom}
\begin{split}
    \pd{}{t}F^{(n)}_{\{\alpha_{i}\};\{\beta_{i}\}}(\{k_{i}\},t) & +  H^{(n)}_{\{\alpha_{i}\};\{\gamma_{i}\}}(\{k_{i}\}) \left( S^{(n)} \right)^{-1}_{\{\gamma_{i}\};\{\delta_{i}\}}(\{k_{i}\})\  F^{(n)}_{\{\delta_{i}\};\{\beta_{i}\}}(\{k_{i}\},t) \\[3pt]
    & + \int_{0}^{t}d\tau M^{(n)}_{\{\alpha_{i}\};\{\gamma_{i}\}}(\{k_{i}\},\tau) \left( H^{(n)} \right)^{-1}_{\{\gamma_{i}\};\{\delta_{i}\}}(\{k_{i}\})\  \pd{}{t}F^{(n)}_{\{\delta_{i}\};\{\beta_{i}\}}(\{k_{i}\},t-\tau) = 0.
\end{split}
\end{equation}
Here, $\{\alpha_{i}\} = \{\alpha_{1},...,\alpha_{n}\}$ and $\{k_{i}\}= \{k_{1},...,k_{n}\}$ denote sets of $n$ particle labels and wave vectors respectively,  and the generalized static structure and collective diffusion tensors are given by (with inverses defined in the same manner as \cref{inv_def}) 
\begin{equation}
    \begin{split}
    & S^{(n)}_{\{\alpha_{i}\};\{\beta_{i}\}}(\{k_{i}\}) = \left\langle \rho^{*}_{\alpha_{1}}(\mv{k}_{1})\hdots \rho^{*}_{\alpha_{n}}(\mv{k}_{n})\rho_{\beta_{1}}(\mv{k}_{1})\hdots \rho_{\beta_{n}}(\mv{k}_{n}) \right\rangle \approx \prod_{i=1}^{n}S_{\alpha_{i}\beta_{i}}(k_{i}),\  \\
    & H^{(n)}_{\{\alpha_{i}\};\{\beta_{i}\}}(\{k_{i}\}) = -\left\langle \rho^{*}_{\alpha_{1}}(\mv{k}_{1})\hdots \rho^{*}_{\alpha_{n}}(\mv{k}_{n})\Omega\rho_{\beta_{1}}(\mv{k}_{1})\hdots \rho_{\beta_{n}}(\mv{k}_{n}) \right\rangle = \sum_{i=1}^{n} \left(D_{\alpha_{i}}k_{i}^{2}\delta_{\alpha_{i}\beta_{i}} \prod_{j\neq i}S_{\alpha_{j}\beta_{j}}(k_{j})  \right),
    \end{split}
\end{equation}
respectively. Finally, using $\{\alpha_{i\neq j}\}$ to represent the set $\{\alpha_{i}\}$ without the element $\alpha_{j}$, introducing the component-dependent number density $n_{\alpha}=N_{\alpha}/V$, and taking the thermodynamic limit, i.e.\ $\sum_{\mv{q}}\rightarrow \frac{V}{(2\pi)^{3}}\int d\mv{q}$, the generalized memory kernel may be written as
\begin{equation}\label{Mnfinal}
    M^{(n)}_{\{\alpha_{i}\};\{\beta_{i}\}}(\{k_{i}\},t) = \int \frac{d\mv{q}}{16\pi^{3}}
    \left(\sum_{j=1}^{n} \frac{D^{\alpha_{j}}}{\sqrt{n_{\alpha_{j}}}} V^{\alpha_{j}}_{\mu \nu}(\mv{k}_{j},\mv{q})F^{(n+1)}_{\mu\nu\{\alpha_{i\neq j}\};\mu^{\prime}\nu^{\prime}\{\beta_{i\neq j}\}}(q,\abs{\mv{k}_{j}-\mv{q}},\{k_{i\neq j}\},t) \frac{D^{\beta_{j}}}{\sqrt{n_{\beta_{j}}}} V^{\beta_{j}}_{\mu^{\prime} \nu^{\prime}}(\mv{k}_{j},\mv{q})  \right). 
\end{equation}
Note that $n=1,2,3$ correspond to \cref{M1final,M2final,M3final} respectively, and that the memory kernel forms the link between each equation of motion and the next via its dependence on the $2(n+1)$-density correlation functions. Overall, we are now left with a hierarchy of connected integro-differential equations which are subject to the initial boundary conditions $F^{(n)}_{\{\alpha_{i}\};\{\beta_{i}\}}(\{k_{i}\},t=0)=S^{(n)}_{\{\alpha_{i}\};\{\beta_{i}\}}(\{k_{i}\})$ and, in principle, can go up to arbitrary order in $n$. For computational reasons however, one must close the hierarchy at a suitable order $n=n_{\mathrm{max}}$. Increasing the value of $n_{\mathrm{max}}$ pushes the factorization to higher order correlation functions (see \cref{numerics} for more details).

To summarize, we have demonstrated that a multi-component GMCT formalism can be constructed for Brownian systems governed by a many-body Smoluchowski equation. A comparison of our presented results with recent work on multi-component Newtonian GMCT~\cite{ciarella2021} shows that in the overdamped limit both hierarchies of equations are completely identical. This demonstrates that the non-trivial similarity between Brownian and Newtonian systems witnessed for standard MCT is maintained for GMCT.

\subsection*{Long-Time Limit}
\noindent We finalize our discussion of the theory by mentioning that, based on the derived equations of motion, we can also formulate a relation for the long-time limit of the $2n$-density correlation functions or the so-called non-ergodicity parameters $f$. Taking the Laplace transform of \cref{Fn_eom} and invoking the final value theorem, we have 
\begin{equation}\label{nep_eom}
\begin{split}
    H^{(n)}_{\{\alpha_{i}\};\{\gamma_{i}\}}(\{k_{i}\}) & \left( S^{(n)} \right)^{-1}_{\{\gamma_{i}\};\{\delta_{i}\}}(\{k_{i}\})\  f^{(n)}_{\{\delta_{i}\};\{\beta_{i}\}}(\{k_{i}\})
     \\[3pt]
    &+   m^{(n)}_{\{\alpha_{i}\};\{\gamma_{i}\}}(\{k_{i}\}) \left( H^{(n)} \right)^{-1}_{\{\gamma_{i}\};\{\delta_{i}\}}(\{k_{i}\})\  \left( f^{(n)}_{\{\delta_{i}\};\{\beta_{i}\}}(\{k_{i}\}) - S^{(n)}_{\{\delta_{i}\};\{\beta_{i}\}}(\{k_{i}\}) \right) = 0,
\end{split}
\end{equation}
where the long-time limits are defined as
\begin{equation}
    f^{(n)}_{\{\alpha_{i}\};\{\beta_{i}\}}(\{k_{i}\}) = \lim_{t\to\infty} F^{(n)}_{\{\alpha_{i}\};\{\beta_{i}\}}(\{k_{i}\},t),
\end{equation}
and
\begin{equation}\label{Mnep}
\begin{split}
    & m^{(n)}_{\{\alpha_{i}\};\{\beta_{i}\}}(\{k_{i}\}) = \lim_{t\to\infty} M^{(n)}_{\{\alpha_{i}\};\{\beta_{i}\}}(\{k_{i}\},t) \\[3pt]
    & \qquad \qquad \qquad \quad = \int \frac{d\mv{q}}{16\pi^{3}}
    \left(\sum_{j=1}^{n} \frac{D^{\alpha_{j}}}{\sqrt{n_{\alpha_{j}}}} V^{\alpha_{j}}_{\mu \nu}(\mv{k}_{j},\mv{q})f^{(n+1)}_{\mu\nu\{\alpha_{i\neq j}\};\mu^{\prime}\nu^{\prime}\{\beta_{i\neq j}\}}(q,\abs{\mv{k}_{j}-\mv{q}},\{k_{i\neq j}\}) \frac{D^{\beta_{j}}}{\sqrt{n_{\beta_{j}}}} V^{\beta_{j}}_{\mu^{\prime} \nu^{\prime}}(\mv{k}_{j},\mv{q})  \right). 
\end{split}
\end{equation}
For the non-ergodicity parameters we thus find a similar hierarchy of, in this case, geometric equations, which must also be closed at a suitable order $n_{\mathrm{max}}$ to allow for the obtainment of practical results. 

\section{Numerical Details}\label{numerics}
\subsection*{GMCT Numerics}
\noindent Our aim is to self-consistently solve the hierarchies of equations for different closure levels $n_{\mathrm{max}}$. To attain such solutions, we will employ the following mean-field closure for $n_{\mathrm{max}}>1$~\cite{Janssen2015a,ciarella2021},
\begin{equation}
    M^{(n_{\mathrm{max}})}_{\{\alpha_{i}\};\{\beta_{i}\}}(\{k_{i}\},t) \approx \frac{1}{n_{\mathrm{max}}-1} \sum_{j} M^{(n_{\mathrm{max}}-1)}_{\{\alpha_{i\neq j}\};\{\beta_{i\neq j}\}}(\{k_{i\neq j}\},t) F_{\alpha_{j};\beta_{j}}(k_{j},t),
\end{equation}
where we have taken into account the invariance of $F^{(n)}_{\{\alpha_{i}\};\{\beta_{i}\}}(\{k_{i}\},t)$ under the exchange $\{ k_{i},\alpha_{i},\beta_{i} \leftrightarrow k_{j},\alpha_{j},\beta_{j} \}$ and the factor $1/(n_{\mathrm{max}}-1)$ ensures that at $t=0$ both sides are equal under Gaussian factorization. Note that this closure is equivalent to assuming $F^{(n_{\mathrm{max}}+1)}(t)\sim F^{(n_{\mathrm{max}})}(t)\times F^{(1)}(t)$ and is therefore consistent with the standard MCT approximation, i.e.\  \cref{MCT_approx}, when setting $n_{\mathrm{max}}=1$. Using these closures, we find explicit expressions for all correlation functions up to $F^{(n_{\mathrm{max}})}_{\{\alpha_{i}\};\{\beta_{i}\}}(\{k_{i}\},t)$, although we will concentrate mainly on $F^{(1)}$, which is usually measured in simulations and experiments. Moreover, by increasing the hierarchy level $n_\mathrm{max}$, the factorization approximation naturally gets shifted towards higher order correlations $F^{(n_{\mathrm{max}}+1)}_{\{\alpha_{i}\};\{\beta_{i}\}}(\{k_{i}\},t)$, which can systematically improve predictions for both single-component and binary systems~\cite{SzamelPRL2003,Wu2005,JanssenPRE2014,Janssen2015a,Janssen2016a,Luo2020_1,Luo2020_2,ciarella2021}. 

Fixing the closure tells us how many integro-differential equations to include, but we also require a numerical scheme to solve each of them. For this we exploit the system's rotational symmetry to rewrite the three-dimensional integral over $\mv{q}$ present in \cref{Mnfinal} in terms of two bipolar coordinates $q=\abs{\mv{q}}$, $p=\abs{\mv{k}_{j}-\mv{q}}$, which are subsequently approximated by a double Riemann sum~\cite{Franosch1997} using an equidistant wavevector grid $k\sigma=[0.2, 0.6,\hdots,39.8]$ (with $\sigma$ the unit of length). To retrieve solutions for the time-evolution equation [\cref{Fn_eom}], we can then utilize a generalization of Fuchs' algorithm~\cite{Fuchs1991}. For this, we determine the first $N_{t}=32$ points in time using a Taylor expansion with a step size $\Delta t=10^{-6}$, subsequently double the time step and numerically integrate the equations of motion, and repeat the duplication each time the next $N_{t}/2$ time points have been calculated.     

\subsection*{Brownian Dynamics Simulations}
\noindent As a proof of principle for our theory we seek to predict the glassy behavior of a Kob-Andersen binary Lennard-Jones (LJ) mixture~\cite{Kob1994} of Brownian particles. This system consists of $N_{\mathrm{A}}=800$, $N_{\mathrm{B}}=200$ particles of type A and B respectively. The position of each particle obeys \cref{eomBrownian}, where we set $\zeta_{\mathrm{A,B}}=\zeta_{0}=1.0$ (so that $D^{\mathrm{A}}=D^{\mathrm{B}}=k_{B}T/\zeta_{0}$ with $k_{B}$ Boltzmann's constant and $T$ the temperature) and the interaction forces are derived from the following inter-particle potential,  
\begin{equation}
    V_{\alpha\beta}(r)=
    \begin{cases}
    4\epsilon_{\alpha\beta}\left[\left(\frac{\sigma_{\alpha\beta}}{r}\right)^{12}-\left(\frac{\sigma_{\alpha\beta}}{r}\right)^{6} +C_{\alpha\beta}\right]~,\qquad r\le r_{\alpha\beta}^c~,
    \\
    \qquad\qquad\qquad 0~,\qquad\qquad\qquad\qquad\quad \ r>r_{\alpha\beta}^c~.
    \end{cases}
\end{equation}
The interaction parameters, i.e.\ $\epsilon_{\mathrm{AA}}=1,\  \epsilon_{\mathrm{AB}}=1.5,\  \epsilon_{\mathrm{BB}}=0.5,\  \sigma_{\mathrm{AA}}=1,\ \sigma_{\mathrm{AB}}=0.8,\  \sigma_{\mathrm{BB}}=0.88$, are chosen to give good glass-forming mixtures and the constant $C_{\alpha\beta}$ fixes the potential to zero at the cutoff radius $r^c_{\alpha\beta}=2.5\sigma_{\alpha\beta}$. Brownian dynamics simulations are then carried out using LAMMPS~\cite{Lammps}. We impose periodic boundary conditions, fix the cubic box size to $L=9.41\sigma_{\mathrm{AA}}$ so that the number density equals $N/V=1.2$, equilibrate the system at different temperatures, and afterwards track the particles over time. All results are presented in reduced units where $\sigma_{\mathrm{AA}}$, $\epsilon_{\mathrm{AA}}$, $\epsilon_{\mathrm{AA}}/k_{\mathrm{B}}$, and $\zeta_{0}\sigma^{2}_{\mathrm{AA}}/\epsilon_{\mathrm{AA}}$ represent the units of length, energy, temperature, and time respectively~\cite{Flenner2005}.

Based on the simulation data, partial structure factors $S_{\alpha ;\beta}(k)$ have been calculated for different temperatures up to two decimal numbers (values at more detailed temperatures are obtained via linear interpolation) and are rewritten in terms of an equidistant grid via cubic spline and polynomial extrapolation for the first two grid points. These structure factors, in combination with the set temperature (or diffusion coefficient) and number densities, then serve as input for the multi-component GMCT equations from which we find the corresponding predicted $F_{\alpha ;\beta}(k,t)$ and $f_{\alpha ;\beta}(k)$. In the following section we will discuss  the effect of increasing the closure level $n_{\mathrm{max}}$ on the predicted dynamics, with a prime focus on the critical temperature, and we also compare our results to previous (Brownian and Newtonian dynamics) simulations of the same system shown in~\cite{Flenner2005,Berthier2010,ciarella2021,Nauroth1997}.   

\section{Results \& Discussion}

\subsection*{Non-Ergodicity Parameter}
\begin{figure}[h!]
    \centering
    \includegraphics[width=0.9\textwidth]{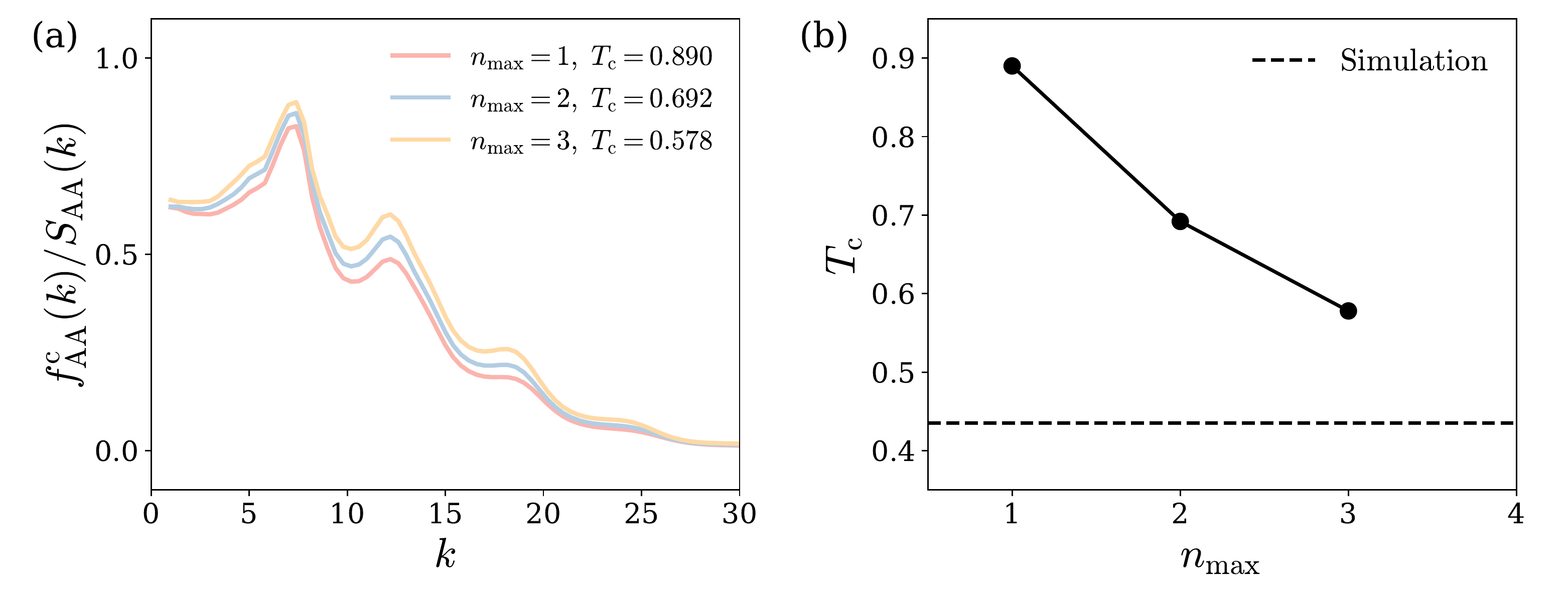}
    \caption{(a) The $\mathrm{AA}$-component (where A is the majority species of the system) of the normalized critical non-ergodicity parameters as a function of wavenumber $k$, calculated for different GMCT closures $n_{\mathrm{max}}$ at the respective critical temperatures $T_{\mathrm{c}}$. (b) The predicted critical temperature $T_c$ as a function of closure level $n_{\mathrm{max}}$. Note that $T_c$ decreases with increasing GMCT closure and seems to systematically move (and possibly converge) towards the value obtained in simulations, $T_{\mathrm{c}}^{\mathrm{sim}}=0.435$ (dashed line).
    }
    \label{fig_nep}
\end{figure}

\noindent Before proceeding to calculate the full dynamics, let us first focus on the long-time limit of the dynamic structure factor $f_{\alpha ; \beta}(k)$. These non-ergodicity parameters serve as a convenient probe to find the critical temperature $T_{\mathrm{c}}$ at which an idealized glass transition occurs according to our theory. During such a transition the value of $f_{\alpha ; \beta}(k)$ jumps discontinuously from zero to a finite value, indicating that the system never fully relaxes and  ergodicity is broken. Solving \cref{nep_eom,Mnep}, and inspecting the resulting non-ergodicity parameters, allows us to retrieve the critical temperature, whose values (up to three decimal points) are listed for different closure levels in \cref{fig_nep}a. It can be seen that $T_{\mathrm{c}}$ decreases in value upon going to higher order GMCT. In fact, when we plot the critical temperature as a function of $n_{\mathrm{max}}$ (see \cref{fig_nep}b), we may observe it moving and possibly converging towards the value $T_{\mathrm{c}}^{\mathrm{sim}}=0.435$ obtained in both Newtonian and Brownian dynamics simulations~\cite{Flenner2005,Berthier2010,Nauroth1997}.   
To confirm a rigorous pattern of convergence, however, we must push our results towards larger values of $n_{\mathrm{max}}$, which so far has proved to be numerically demanding and is therefore left for future work. 

Having specified the critical points, we now  take a closer look at the details of the corresponding non-ergodicity parameters. For convenience, we have only plotted the AA-component in \cref{fig_nep}a, since type-A particles comprise $80$ percent of the system and thus dominate its dynamics. The curves show that the non-ergodicity parameter (normalized by the static structure factor) at the critical temperature retains a qualitatively similar shape for all considered closure levels. In particular, this shape closely resembles previously obtained results of a single-component Percus-Yevick hard sphere system~\cite{Luo2020_1}. The quantitative increase for larger $n_{\mathrm{max}}$ is also consistent with the results of Ref.~\cite{Luo2020_1} and physically reflects a slowdown of the collective relaxation dynamics. To rigorously test whether the increase of the normalized non-ergodicity parameter proceeds in a convergent manner towards simulation results~\cite{Nauroth1997}, akin to the observed decrease of the critical temperature, will require higher orders $n_{\mathrm{max}}$ to be included and a more accurate pinning down of the critical temperatures~\cite{Luo2020_1}; due to computational complexity, the critical temperature is here only determined up to three decimal numbers. Still, these results already suggest that extending the framework of GMCT from monodisperse towards multi-component  systems keeps its improved predictive power intact and does not introduce any new qualitative flaws.

\subsection*{Dynamics and the Relaxation Time}
\begin{figure}[ht!]
    \centering
    \includegraphics[width=1.0\textwidth]{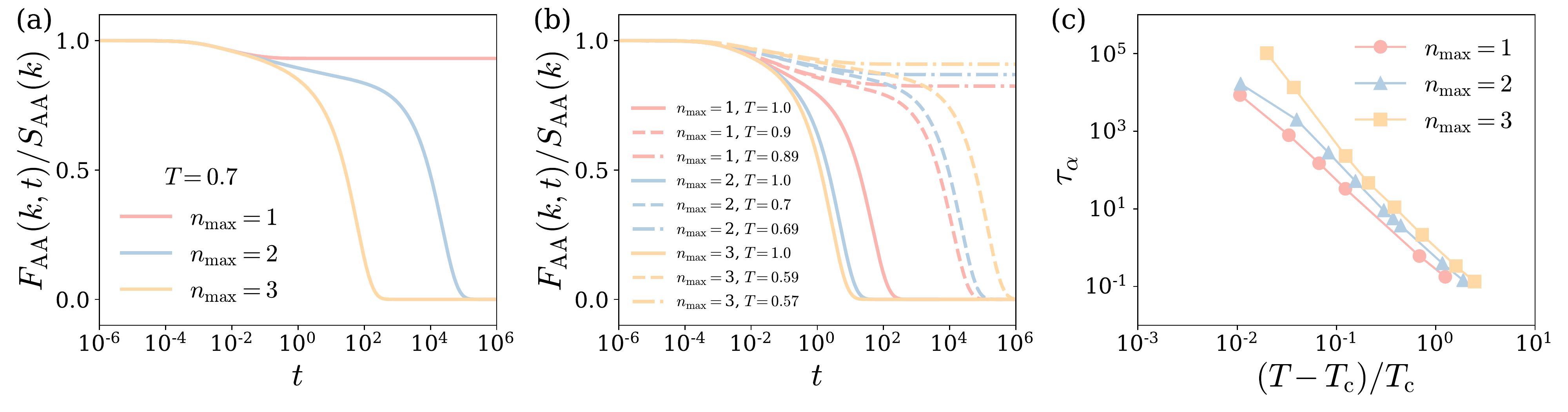}
    \caption{(a-b) Intermediate scattering functions as a function of time obtained from binary GMCT. The results correspond to (a) different closure levels at a fixed temperature $T=0.7$ and (b) different closure levels and different temperatures. We present only the AA-component, since type A species dominate the dynamics, and have chosen a wavenumber $k=7.4\sigma_{\mathrm{AA}}^{-1}$ close to the first peak of $S_{\mathrm{AA}}(k)$. Panel (c) shows the structural relaxation time (obtained for the same wavenumber) as a function of the reduced temperature for different GMCT closure levels.
    }
    \label{fig_Ft}
\end{figure}

\noindent Next, to study the effect of going to higher order GMCT on the full dynamics of the system, we have also solved the time-dependent hierarchy of equations posed by \cref{Fn_eom,Mnfinal} for different closure levels $n_{\mathrm{max}}$ and temperatures $T$. The calculated dynamic structure factors (again focusing only on the AA-component and normalized by the static structure factor) are shown for the first three closure levels and a subset of considered temperatures in \cref{fig_Ft}a-b. To illustrate the quantitatively improved predictive power of GMCT, we first focus on \cref{fig_Ft}a where the temperature is kept fixed at a value of $T=0.7$ and only the closure level has been varied. Being relatively far from the critical temperature $T_{\mathrm{c}}^{\mathrm{sim}}=0.435$, one expects a liquid state with the 
%dynamic structure factor 
intermediate scattering function
relaxing to zero at a timescale on the order of $\sim 10^{1}$~\cite{Flenner2005}. However, at this temperature standard MCT ($n_{\mathrm{max}}=1$) incorrectly predicts a non-ergodic glassy state, evident from the fact that $F_{\mathrm{AA}}(k,t)$ retains a plateau value and does not relax to zero. In comparison, second order GMCT ($n_{\mathrm{max}}=2$) already correctly predicts a liquid state, although the time scale at which the %partial structure factor
$2$-point density correlation function
decays to zero is still overestimated with respect to the values obtained in simulations. This timescale is seen to be pushed back further towards its expected value upon moving to third order GMCT ($n_{\mathrm{max}}=3$), thereby demonstrating how GMCT can systematically improve its predictions and possibly approach actual simulation results in the limit $n_{\mathrm{max}}\rightarrow \infty$. 

We now proceed to the temperature dependence of the intermediate scattering functions, shown in  \cref{fig_Ft}b. For all closure levels, it is apparent that the relaxation of $F_{\mathrm{AA}}(k,t)$ takes an increasingly longer time upon cooling the system, until, at a small enough temperature, it fails to relax over the entire simulated time range. An inspection of the presented temperatures shows that the points at which the intermediate scattering functions cease to decay to zero are in agreement with the critical points obtained from the long-time limit calculations. Additionally, we may observe that the overall shape of the relaxation curves is similar across all closure levels. This suggests that higher order GMCT does not yield strong qualitative changes and we expect the celebrated MCT scaling laws~\cite{Franosch1997,gotze2008complex,reichman2005mode} to also hold for multi-component GMCT. Note that this has already been rigorously shown for single component GMCT~\cite{Luo2020_1,Luo2020_2}. 
%\begin{figure}[ht!]
%    \centering
%    \includegraphics[width=0.75\textwidth]{Figure_2_2.pdf}
%    \caption{(a-c) Plots of the normalized dynamic structure factor as a function of time obtained from binary GMCT. The results correspond to different temperatures and closure levels. We present only the AA-component, since type A species dominate the dynamics, and have chosen a wavenumber $k=7.4\sigma_{\mathrm{AA}}^{-1}$ close to the first peak of $S_{\mathrm{AA}}(k)$. (d) The relaxation time (obtained for the same wavenumber) is shown as function of the reduced temperature for different closure levels.
%    }
%    \label{fig_Ft}
%\end{figure}

We conclude our discussion by exploring one of such scaling laws, namely the power law divergence of the relaxation time, for our binary LJ mixture. As an operational definition we follow~\cite{Flenner2005,Berthier2010} and let the relaxation time $\tau_{\alpha}$ mark the point at which the intermediate scattering function has decreased to $e^{-1}$ of its initial value, i.e.\ $F_{\mathrm{AA}}(k,\tau_{\alpha})/S_{\mathrm{AA}}(k)=e^{-1}$ (with the wavenumber $k$ corresponding to the first peak of $S_{\mathrm{AA}}(k)$). When described as a function of the reduced temperature $\epsilon=(T-T_{\mathrm{c}})/T_{\mathrm{c}}$, these relaxation times are expected to follow a power law divergence given by $\tau_{\alpha}\propto \epsilon^{-\gamma}$, with $\gamma$ a critical exponent related to the fragility of the material. We have plotted the resulting values of $\tau_{\alpha}$ for different closures in \cref{fig_Ft}. It can be seen that they initially (down to $\epsilon \gtrsim  0.1$) follow a power law (represented by a straight line) before variations start to occur for the higher order GMCT results. We believe that the observed deviations from a straight line when approaching the critical temperature (smaller $\epsilon$) to be a result of subtle inaccuracies in the static structure factors obtained from simulations. Since changes in the static structure factor are amplified for increasing order $n_{\mathrm{max}}$, this can explain why deviations become more evident when we go beyond MCT, while for the latter an approximately straight line is retained over the full investigated window of reduced temperatures. Alternatively, it is also possible that our results are not fully converged yet with respect to the wavenumber grid, which for computational reasons we have limited to $N_{k}=100$ $k$-points. Taking this into account, we have only fitted the relaxation times up to a value of $\epsilon \sim 0.1$,  yielding for all closures an exponent between $\gamma\approx 2.3-2.4$. These values are in good agreement with the ones retrieved in Refs.~\cite{Flenner2005,Berthier2010} and suggests that increasing the GMCT closure does not dramatically alter the critical exponent and thus the calculated fragility for this particular model system.

\section{Conclusion}
\noindent In this work we have presented the first derivation of a fully time-dependent, microscopic, generalized MCT for colloidal mixtures obeying a many-body Smoluchowski equation.
Our theory is inspired by earlier, time-independent GMCT studies on monodisperse colloidal systems  MCT~\cite{SzamelPRL2003,Wu2005} and by previous work on time-dependent Newtonian GMCT~\cite{Janssen2015a,Janssen2016a,ciarella2021}. 
The framework we put forward consists of a hierarchy of coupled integro-differential equations describing the time-dependence of diagonal density correlation functions of increasing order. For the first three entries, i.e.\ the $2$-, $4$-, and $6$-point density correlation functions, the corresponding equations of motion (and their dependence on the $4$-, $6$-, and $8$-point density correlation functions, respectively) have been explicitly calculated. Subsequently, we have extrapolated this hierarchical construct to describe correlation functions of arbitrary order and thereby established our fully generalized MCT. Using only the partial static structure factors and no additional fit parameters as input, the hierarchy can be closed at an order $n$ and subsequently solved self-consistently to retrieve the relaxation dynamics of a colloidal mixture and study its glassy behavior. Moreover, a careful inspection of the involved equations has shown that they, consistent with standard MCT, are \textit{identical} to the ones obtained from Newtonian GMCT~\cite{ciarella2021} when taking the overdamped limit. The non-trivial similarity between Brownian and Newtonian MCT is therefore maintained for our newly developed multi-component GMCT, where we highlight that such an equivalency is also expected based on simulation results~\cite{Flenner2005,Berthier2010,ciarella2021,Nauroth1997}.

As an explorative demonstration of the theory, we have used it to predict the glassy behavior of a three-dimensional Kob-Andersen binary LJ mixture of Brownian particles. For such a system the theory yields an ergodicity-breaking (idealized glass) transition at decreasing critical temperatures upon increasing the closure level $n_{\mathrm{max}}$. In particular, this decrease seems to occur (at least for the considered closure levels) in a convergent manner towards the actual experimental value obtained in both Newtonian and Brownian dynamics simulations. Additionally, a careful study of the calculated non-ergodicity parameters and intermediate scattering functions suggests that no strong qualitative changes occur upon increasing the closure level, while quantitative improvements are observed. These results support the role of GMCT as a promising, systematically improvable, first-principles theory to study the dynamics of glass-forming materials.

Finally, we want to mention the recent heightened interest in active glassy materials where, besides exhibiting passive motion, the constituent particles are also able to autonomously move through the consumption of energy \cite{Janssen2019active}. For such materials several active MCT theories have recently been put forward~\cite{Voigtmann2017active,SzamelABP2019,SzamelAOUP2016,FengHou2017,FengHou2018,Reichert2021,farage2014dynamics}. Since most model active systems are usually considered in the overdamped limit, e.g.\ active Brownian particles, our colloidal GMCT framework might also serve as a stepping stone for the development of active GMCT theories.

\section*{Acknowledgments}
\noindent We acknowledge the Dutch Research Council (NWO) for financial support through a START-UP grant (V.E.D., C.L., and L.M.C.J.) and Vidi grant (L.M.C.J.).

%\section*{SUPPLEMENTARY MATERIAL}

\bibliographystyle{apsrev4-1}
\bibliography{all}

% Figure legends
%%Automatically it will add the figure legends  and table legends as a list by below command

\newpage

\section*{Supplementary Information}

\subsection*{Mori-Zwanzig Formalism}
\noindent In standard MCT one applies the Mori-Zwanzig formalism~\cite{Mori65,Zwanzig60} to derive an equation of motion for the $2$-point density correlation functions (or intermediate scattering functions), which can be subsequently solved by using factorization approximations. We now seek to show that the formalism can be equally applied to higher order (diagonal) correlation functions. In particular, we will demonstrate this for the $4$- and $6$-point density correlation functions. Before starting, we reiterate that the correlation functions are described in terms of density modes (the Fourier components of the microscopic particle densities), 
\begin{equation}
    \rho_{\alpha} (\mv{k})=\frac{1}{\sqrt{N_{\alpha}}}\sum_{j=1}^{N_{\alpha}} e^{i\mv{k}\cdot \mv{r}^{\alpha}_{j}},
\end{equation}
with wavevector $\mv{k}$ and $N_{\alpha}$ the number of particles of type $\alpha$.
The time-dependence is imposed via the evolution (or Smoluchowski) operator,
\begin{equation}
    \Omega = \sum_{\alpha=1}^{m}\sum_{i=1}^{N_{\alpha}}  D^{\alpha}_{t} \nabla^{\alpha}_{i} \cdot \left( \nabla^{\alpha}_{i} - \beta  \mv{F}^{\alpha}_{i}  \right),
\end{equation}
where $\beta$ is the inverse thermal energy, $D^{\alpha}$ the diffusion coefficient corresponding to type $\alpha$, and $\mv{F}^{\alpha}_{i}$ the force acting on particle $i$. Moreover, averages $\avg{\hdots}$ are taken with respect to a Boltzmann distribution function $P_{\mathrm{eq}}(\mv{r}^{N})\propto \exp\left(-\beta U(\mv{r}^{N})\right)$, which depends solely on the total (spherically symmetric) interaction potential $U(\mv{r}^{N})$ from which the interaction forces are derived, i.e.\ $\mv{F}^{\alpha}_{i}=-\nabla^{\alpha}_{i}U(\mv{r}^{N})$.

\subsubsection*{Four-point density correlation function}
\noindent Starting with the $4$-point density correlation function, we have as its definition
\begin{equation}
F^{(2)}_{\alpha_{1}\alpha_{2};\beta_{1}\beta_{2}}(k_{1},k_{2},t) =\avg{\rho_{\alpha_{1}}^{*} (\mv{k}_{1})\rho_{\alpha_{2}}^{*} (\mv{k}_{2}) e^{\Omega t} \rho_{\beta_{1}} (\mv{k}_{1})\rho_{\beta_{2}} (\mv{k}_{2})},
\end{equation}
and, for convenience, take its Laplace transform ($\mathcal{LT}[\hdots]$)
\begin{equation}
\mathcal{LT} [F^{(2)}_{\alpha_{1}\alpha_{2};\beta_{1}\beta_{2}}(k_{1},k_{2},t)] \equiv F^{(2)}_{\alpha_{1}\alpha_{2};\beta_{1}\beta_{2}}(k_{1},k_{2},z) =\avg{\rho_{\alpha_{1}}^{*} (\mv{k}_{1})\rho_{\alpha_{2}}^{*} (\mv{k}_{2}) (z-\Omega)^{-1} \rho_{\beta_{1}} (\mv{k}_{1})\rho_{\beta_{2}} (\mv{k}_{2})}.
\end{equation}
Now we can invoke Dyson decomposition, i.e.\ 
\begin{equation}
    (z-\Omega)^{-1} = (z-\Omega\mathcal{Q}^{(2)})^{-1} + (z-\Omega\mathcal{Q}^{(2)})^{-1} \Omega \mathcal{P}^{(2)}  (z-\Omega)^{-1},
\end{equation}
with the projection operator (again we adopt the Einstein summation convention to sum over repeated indices throughout)
\begin{equation}
\mathcal{P}^{(2)}=  \rho_{\alpha_{1}} (\mv{k}_{1})\rho_{\alpha_{2}} (\mv{k}_{2}) \rangle  \left(S^{(2)}\right)^{-1}_{\alpha_{1}\alpha_{2};\beta_{1}\beta_{2}}(k_{1},k_{2})\ 
\langle \rho_{\beta_{1}}^{*} (\mv{k}_{1})\rho_{\beta_{2}}^{*} (\mv{k}_{2}),
\end{equation}
and its orthogonal complement $\mathcal{Q}^{(2)}=\mathcal{I}-\mathcal{P}^{(2)}$, to rewrite the Laplace transform of its time-derivative as
\begin{equation}\label{F2eom}
\begin{split}
     \hspace{-6.0cm} \mathcal{LT}[\dot{F}^{(2)}_{\alpha_{1}\alpha_{2};\beta_{1}\beta_{2}}(k_{1},k_{2},t)] & = z F^{(2)}_{\alpha_{1}\alpha_{2};\beta_{1}\beta_{2}}(k_{1},k_{2},z) - S^{(2)}_{\alpha_{1}\alpha_{2};\beta_{1}\beta_{2}}(k_{1},k_{2}) \\[4pt]
    & = \avg{\rho_{\alpha_{1}}^{*} (\mv{k}_{1})\rho_{\alpha_{2}}^{*} (\mv{k}_{2}) \Omega  (\mathcal{P}^{(2)} + \mathcal{Q}^{(2)}) (z-\Omega)^{-1}\rho_{\beta_{1}} (\mv{k}_{1})\rho_{\beta_{2}} (\mv{k}_{2})} \\[4pt]
    & = -H^{(2)}_{\alpha_{1}\alpha_{2};\gamma_{1}\gamma_{2}}(k_{1},k_{2}) \left(S^{(2)}\right)^{-1}_{\gamma_{1}\gamma_{2};\epsilon_{1}\epsilon_{2}}(k_{1},k_{2})\  F^{(2)}_{\epsilon_{1}\epsilon_{2};\beta_{1}\beta_{2}}(k_{1},k_{2},z) \\[4pt]
    & \hspace{0.4cm} +  K^{(2)}_{\alpha_{1}\alpha_{2};\gamma_{1}\gamma_{2}}(k_{1},k_{2},z) \left(S^{(2)}\right)^{-1}_{\gamma_{1}\gamma_{2};\epsilon_{1}\epsilon_{2}}(k_{1},k_{2})\  F^{(2)}_{\epsilon_{1}\epsilon_{2};\beta_{1}\beta_{2}}(k_{1},k_{2},z).
\end{split}
\end{equation}
Here, we have introduced the $4$-point static density correlation
\begin{equation}
    S^{(2)}_{\alpha_{1}\alpha_{2};\beta_{1}\beta_{2}}(k_{1},k_{2})=\avg{\rho_{\alpha_{1}}^{*} (\mv{k}_{1})\rho_{\alpha_{2}}^{*} (\mv{k}_{2}) \rho_{\beta_{1}} (\mv{k}_{1})\rho_{\beta_{2}} (\mv{k}_{2})},
\end{equation}
the collective $4$-point diffusion tensor
\begin{equation}
    H^{(2)}_{\alpha_{1}\alpha_{2};\beta_{1}\beta_{2}}(k_{1},k_{2})= - \avg{\rho_{\alpha_{1}}^{*} (\mv{k}_{1})\rho_{\alpha_{2}}^{*} (\mv{k}_{2})\Omega \rho_{\beta_{1}} (\mv{k}_{1})\rho_{\beta_{2}} (\mv{k}_{2})}=k_{1}^{2}D^{\alpha_{1}}\delta_{\alpha_{1}\beta_{1}}S_{\alpha_{2};\beta_{2}}(k_{2})+k_{2}^{2}D^{\alpha_{2}}\delta_{\alpha_{2}\beta_{2}}S_{\alpha_{1};\beta_{1}}(k_{1}),
\end{equation}
and the memory kernel
\begin{equation}
    K^{(2)}_{\alpha_{1}\alpha_{2};\beta_{1}\beta_{2}}(k_{1},k_{2},z)=\avg{\rho_{\alpha_{1}}^{*} (\mv{k}_{1})\rho_{\alpha_{2}}^{*} (\mv{k}_{2}) \Omega \mathcal{Q}^{(2)} (z-\mathcal{Q}^{(2)}\Omega \mathcal{Q}^{(2)})^{-1} \mathcal{Q}^{(2)} \Omega \rho_{\beta_{1}} (\mv{k}_{1})\rho_{\beta_{2}} (\mv{k}_{2})}.
\end{equation}
Moreover, we mention that inverse tensors (denoted by the superscript $-1$) are used to ensure idempotency of the projection operators and are thus defined according to
\begin{equation}
    S^{(2)}_{\alpha_{1}\alpha_{2};\gamma_{1}\gamma_{2}}(k_{1},k_{2}) \left(S^{(2)}\right)^{-1}_{\gamma_{1}\gamma_{2};\beta_{1}\beta_{2}}(k_{1},k_{2}) = \delta_{\alpha_{1}\beta_{1}}\delta_{\alpha_{2}\beta_{2}}.
\end{equation}
In this form, however, the memory kernel does not lend itself to MCT-like approximations and we therefore follow standard procedure by converting it to an irreducible memory kernel. Employing a further projection operator,
\begin{equation}
\mathcal{P}_{\mathrm{irr}}^{(2)}=  - \rho_{\alpha_{1}} (\mv{k}_{1})\rho_{\alpha_{2}} (\mv{k}_{2}) \rangle \left(H^{(2)}\right)^{-1}_{\alpha_{1}\alpha_{2};\beta_{1}\beta_{2}}(k_{1},k_{2})\ 
\langle \rho_{\beta_{1}}^{*} (\mv{k}_{1})\rho_{\beta_{2}}^{*} (\mv{k}_{2}) \Omega, \hspace{1.0cm} \mathcal{Q}_{\mathrm{irr}}^{(2)}= \mathcal{I} - \mathcal{P}_{\mathrm{irr}}^{(2)},
\end{equation}
and again applying Dyson decomposition, we find that
\begin{equation}\label{Mirr_eom}
    K^{(2)}_{\alpha_{1}\alpha_{2};\beta_{1}\beta_{2}}(k_{1},k_{2},z)=M^{(2)}_{\alpha_{1}\alpha_{2};\beta_{1}\beta_{2}}(k_{1},k_{2},z) - M^{(2)}_{\alpha_{1}\alpha_{2};\gamma_{1}\gamma_{2}}(k_{1},k_{2},z)\left(H^{(2)}\right)^{-1}_{\gamma_{1}\gamma_{2};\epsilon_{1}\epsilon_{2}}(k_{1},k_{2})\ K^{(2)}_{\epsilon_{1}\epsilon_{2};\beta_{1}\beta_{2}}(k_{1},k_{2},z),
\end{equation}
where the irreducible memory kernel yields 
\begin{equation}
    M^{(2)}_{\alpha_{1}\alpha_{2};\beta_{1}\beta_{2}}(k_{1},k_{2},z)=\avg{\rho_{\alpha_{1}}^{*} (\mv{k}_{1})\rho_{\alpha_{2}}^{*} (\mv{k}_{2}) \Omega \mathcal{Q}^{(2)} (z-\Omega^{(2)}_{\mathrm{irr}})^{-1} \mathcal{Q}^{(2)} \Omega \rho_{\beta_{1}} (\mv{k}_{1})\rho_{\beta_{2}} (\mv{k}_{2})},
\end{equation}
and the corresponding irreducible evolution operator is defined as $\Omega^{(2)}_{\mathrm{irr}}=\mathcal{Q}^{(2)}\Omega\mathcal{Q}_{\mathrm{irr}}^{(2)}\mathcal{Q}^{(2)}$. If we then combine \cref{F2eom,Mirr_eom}, we finally obtain
\begin{equation}
    \begin{split}
    & \left[\delta_{\alpha_{1}\epsilon_{1}}\delta_{\alpha_{2}\epsilon_{2}} + M^{(2)}_{\alpha_{1}\alpha_{2};\gamma_{1}\gamma_{2}}(k_{1},k_{2},z)  \left(H^{(2)}\right)^{-1}_{\gamma_{1}\gamma_{2};\epsilon_{1}\epsilon_{2}}(k_{1},k_{2}) \right]  \bigg[ z F^{(2)}_{\epsilon_{1}\epsilon_{2};\beta_{1}\beta_{2}}(k_{1},k_{2},z) - S^{(2)}_{\epsilon_{1}\epsilon_{2};\beta_{1}\beta_{2}}(k_{1},k_{2}) \bigg] \\[4pt] & \hspace{1.75cm} + H^{(2)}_{\alpha_{1}\alpha_{2};\gamma_{1}\gamma_{2}}(k_{1},k_{2}) \left(S^{(2)}\right)^{-1}_{\gamma_{1}\gamma_{2};\epsilon_{1}\epsilon_{2}}(k_{1},k_{2})\  F^{(2)}_{\epsilon_{1}\epsilon_{2};\beta_{1}\beta_{2}}(k_{1},k_{2},z) = 0,
    \end{split}
\end{equation}
which, after taking the inverse Laplace transform, results in the equation of motion presented in the main text:
\begin{equation}
    \begin{split}
    \pd{}{t}F^{(2)}_{\alpha_{1}\alpha_{2};\beta_{1}\beta_{2}}(k_{1},k_{2},t) & + H^{(2)}_{\alpha_{1}\alpha_{2};\gamma_{1}\gamma_{2}}(k_{1},k_{2}) \left(S^{(2)}\right)^{-1}_{\gamma_{1}\gamma_{2};\epsilon_{1}\epsilon_{2}}(k_{1},k_{2})\  F^{(2)}_{\epsilon_{1}\epsilon_{2};\beta_{1}\beta_{2}}(k_{1},k_{2},t)\\
    &+ \int_{0}^{t} dt^{\prime}  M^{(2)}_{\alpha_{1}\alpha_{2};\gamma_{1}\gamma_{2}}(k_{1},k_{2},t-t^{\prime}) \left(H^{(2)}\right)^{-1}_{\gamma_{1}\gamma_{2};\epsilon_{1}\epsilon_{2}}(k_{1},k_{2})\  \pd{}{t^{\prime}}F^{(2)}_{\epsilon_{1}\epsilon_{2};\beta_{1}\beta_{2}}(k_{1},k_{2},t^{\prime}) =0.
    \end{split}
\end{equation}

\subsubsection*{Six-point density correlation function}
\noindent Next, we proceed to the $6$-point density correlation function, which is defined as
\begin{equation}
F^{(3)}_{\alpha_{1}\alpha_{2}\alpha_{3};\beta_{1}\beta_{2}\beta_{3}}(k_{1},k_{2},k_{3},t) =\avg{\rho_{\alpha_{1}}^{*} (\mv{k}_{1})\rho_{\alpha_{2}}^{*} (\mv{k}_{2}) \rho_{\alpha_{3}}^{*} (\mv{k}_{3}) e^{\Omega t} \rho_{\beta_{1}} (\mv{k}_{1})\rho_{\beta_{2}} (\mv{k}_{2})\rho_{\beta_{3}} (\mv{k}_{3})}.
\end{equation}
Following a similar strategy as outlined for the $4$-point density correlation, we may use the projection operators 
\begin{equation}
\mathcal{P}^{(3)}=  \rho_{\alpha_{1}} (\mv{k}_{1})\rho_{\alpha_{2}} (\mv{k}_{2})\rho_{\alpha_{3}} (\mv{k}_{3}) \rangle \left(S^{(3)}\right)^{-1}_{\alpha_{1}\alpha_{2}\alpha_{3};\beta_{1}\beta_{2}\beta_{3}}(k_{1},k_{2},k_{3})\ 
\langle \rho_{\beta_{1}}^{*} (\mv{k}_{1})\rho_{\beta_{2}}^{*} (\mv{k}_{2})\rho_{\beta_{3}}^{*} (\mv{k}_{3}), \hspace{1.0cm} \mathcal{Q}^{(3)}= \mathcal{I} - \mathcal{P}^{(3)},
\end{equation}
to obtain a relation between the $6$-point density correlation function and memory kernel in the Laplace domain, i.e.\ 
\begin{equation}\label{F3eom}
\begin{split}
    & z F^{(3)}_{\alpha_{1}\alpha_{2}\alpha_{3};\beta_{1}\beta_{2}\beta_{3}}(k_{1},k_{2},k_{3},z) - S^{(3)}_{\alpha_{1}\alpha_{2}\alpha_{3};\beta_{1}\beta_{2}\beta_{3}}(k_{1},k_{2},k_{3}) = -H^{(3)}_{\alpha_{1}\alpha_{2}\alpha_{3};\gamma_{1}\gamma_{2}\gamma_{3}}(k_{1},k_{2},k_{3}) \left(S^{(3)}\right)^{-1}_{\gamma_{1}\gamma_{2}\gamma_{3};\epsilon_{1}\epsilon_{2}\epsilon_{3}}(k_{1},k_{2},k_{3})  \\[4pt]
    & \hspace{0.5cm} F^{(3)}_{\epsilon_{1}\epsilon_{2}\epsilon_{3};\beta_{1}\beta_{2}\beta_{3}}(k_{1},k_{2},k_{3},z) +  K^{(3)}_{\alpha_{1}\alpha_{2}\alpha_{3};\gamma_{1}\gamma_{2}\gamma_{3}}(k_{1},k_{2},k_{3}) \left(S^{(3)}\right)^{-1}_{\gamma_{1}\gamma_{2}\gamma_{3};\epsilon_{1}\epsilon_{2}\epsilon_{3}}(k_{1},k_{2},k_{3})\  F^{(3)}_{\epsilon_{1}\epsilon_{2}\epsilon_{3};\beta_{1}\beta_{2}\beta_{3}}(k_{1},k_{2},k_{3},z).
\end{split}
\end{equation}
In this equation we have introduced the $6$-point counterparts of the static density correlation
\begin{equation}
    S^{(3)}_{\alpha_{1}\alpha_{2}\alpha_{3};\beta_{1}\beta_{2}\beta_{3}}(k_{1},k_{2},k_{3})=\avg{\rho_{\alpha_{1}}^{*} (\mv{k}_{1})\rho_{\alpha_{2}}^{*} (\mv{k}_{2}) \rho_{\alpha_{3}}^{*} (\mv{k}_{3}) \rho_{\beta_{1}} (\mv{k}_{1})\rho_{\beta_{2}} (\mv{k}_{2})\rho_{\beta_{3}} (\mv{k}_{3})},
\end{equation}
the collective diffusion tensor
\begin{equation}
\begin{split}
    H^{(3)}_{\alpha_{1}\alpha_{2}\alpha_{3};\beta_{1}\beta_{2}\beta_{3}} (k_{1},k_{2},k_{3}) & = - \avg{\rho_{\alpha_{1}}^{*} (\mv{k}_{1})\rho_{\alpha_{2}}^{*} (\mv{k}_{2}) \rho_{\alpha_{1}}^{*} (\mv{k}_{3})\Omega \rho_{\beta_{1}} (\mv{k}_{1})\rho_{\beta_{2}} (\mv{k}_{2})\rho_{\beta_{3}} (\mv{k}_{3})} \\[4pt]
    & = k_{1}^{2}D^{\alpha_{1}}\delta_{\alpha_{1}\beta_{1}}S^{(2)}_{\alpha_{2}\alpha_{3};\beta_{2}\beta_{3}}(k_{2},k_{3})+k_{2}^{2}D^{\alpha_{2}}\delta_{\alpha_{2}\beta_{2}}S^{(2)}_{\alpha_{1}\alpha_{3};\beta_{1}\beta_{3}}(k_{1},k_{3})+k_{3}^{2}D_{t}^{\alpha_{3}}\delta_{\alpha_{3}\beta_{3}}S^{(2)}_{\alpha_{1}\alpha_{2};\beta_{1}\beta_{2}}(k_{1},k_{2}),
\end{split}
\end{equation}
and the memory kernel
\begin{equation}
    K^{(3)}_{\alpha_{1}\alpha_{2}\alpha_{3};\beta_{1}\beta_{2}\beta_{3}}(k_{1},k_{2},k_{3},z)=\avg{\rho_{\alpha_{1}}^{*} (\mv{k}_{1})\rho_{\alpha_{2}}^{*} (\mv{k}_{2}) \rho_{\alpha_{3}}^{*} (\mv{k}_{3}) \Omega \mathcal{Q}^{(3)} (z-\mathcal{Q}^{(3)}\Omega \mathcal{Q}^{(3)})^{-1} \mathcal{Q}^{(3)} \Omega \rho_{\beta_{1}} (\mv{k}_{1})\rho_{\beta_{2}} (\mv{k}_{2})\rho_{\beta_{3}} (\mv{k}_{3})}.
\end{equation}
An additional projection operator,
\begin{equation}
\mathcal{P}_{\mathrm{irr}}^{(3)}=  - \rho_{\alpha_{1}} (\mv{k}_{1})\rho_{\alpha_{2}} (\mv{k}_{2})\rho_{\alpha_{3}} (\mv{k}_{3}) \rangle \left(H^{(3)}\right)^{-1}_{\alpha_{1}\alpha_{2}\alpha_{3};\beta_{1}\beta_{2}\beta_{3}}(k_{1},k_{2},k_{3})\ 
\langle \rho_{\beta_{1}}^{*} (\mv{k}_{1})\rho_{\beta_{2}}^{*} (\mv{k}_{2}) \rho_{\beta_{3}}^{*} (\mv{k}_{3}) \Omega, \hspace{0.8cm} \mathcal{Q}_{\mathrm{irr}}^{(3)}= \mathcal{I} - \mathcal{P}_{\mathrm{irr}}^{(3)},
\end{equation}
then allows us to rewrite the memory kernel in terms of a irreducible memory kernel via
\begin{equation}\label{Mirr3_eom}
\begin{split}
    K^{(3)}_{\alpha_{1}\alpha_{2}\alpha_{3};\beta_{1}\beta_{2}\beta_{3}}(k_{1},k_{2},k_{3},z) = &\ M^{(3)}_{\alpha_{1}\alpha_{2}\alpha_{3};\beta_{1}\beta_{2}\beta_{3}}(k_{1},k_{2},k_{3},z) - M^{(3)}_{\alpha_{1}\alpha_{2}\alpha_{3};\gamma_{1}\gamma_{2}\gamma_{3}}(k_{1},k_{2},k_{3},z)\\
    & \left(H^{(3)}\right)^{-1}_{\gamma_{1}\gamma_{2}\gamma_{3};\epsilon_{1}\epsilon_{2}\epsilon_{3}}(k_{1},k_{2},k_{3})\ K^{(3)}_{\epsilon_{1}\epsilon_{2}\epsilon_{3};\beta_{1}\beta_{2}\beta_{3}}(k_{1},k_{2},k_{3},z),
\end{split}
\end{equation}
with the irreducible memory kernel given by 
\begin{equation}\label{M3_formal}
    M^{(3)}_{\alpha_{1}\alpha_{2}\alpha_{3};\beta_{1}\beta_{2}\beta_{3}}(k_{1},k_{2},k_{3},z)=\avg{\rho_{\alpha_{1}}^{*} (\mv{k}_{1})\rho_{\alpha_{2}}^{*} (\mv{k}_{2})\rho_{\alpha_{3}}^{*} (\mv{k}_{3}) \Omega \mathcal{Q}^{(3)} (z-\Omega^{(3)}_{\mathrm{irr}})^{-1} \mathcal{Q}^{(3)} \Omega \rho_{\beta_{1}} (\mv{k}_{1})\rho_{\beta_{2}} (\mv{k}_{2})\rho_{\beta_{3}} (\mv{k}_{3})},
\end{equation}
and the corresponding irreducible evolution operator being defined as $\Omega^{(3)}_{\mathrm{irr}}=\mathcal{Q}^{(3)}\Omega\mathcal{Q}_{\mathrm{irr}}^{(3)}\mathcal{Q}^{(3)}$. Based on \cref{F3eom,Mirr3_eom}, we can then derive that
\begin{equation}
    \begin{split}
    & \left[\delta_{\alpha_{1}\epsilon_{1}}\delta_{\alpha_{2}\epsilon_{2}}\delta_{\alpha_{3}\epsilon_{3}} + M^{(3)}_{\alpha_{1}\alpha_{2}\alpha_{3};\gamma_{1}\gamma_{2}\gamma_{3}}(k_{1},k_{2},k_{3},z)  \left(H^{(3)}\right)^{-1}_{\gamma_{1}\gamma_{2}\gamma_{3};\epsilon_{1}\epsilon_{2}\epsilon_{3}}(k_{1},k_{2},k_{3}) \right]  \bigg[ z F^{(3)}_{\epsilon_{1}\epsilon_{2}\epsilon_{3};\beta_{1}\beta_{2}\beta_{3}}(k_{1},k_{2},k_{3},z) \\[4pt] 
    & - S^{(3)}_{\epsilon_{1}\epsilon_{2}\epsilon_{3};\beta_{1}\beta_{2}\beta_{3}}(k_{1},k_{2},k_{3}) \bigg] + H^{(3)}_{\alpha_{1}\alpha_{2}\alpha_{3};\gamma_{1}\gamma_{2}\gamma_{3}}(k_{1},k_{2},k_{3}) \left(S^{(3)}\right)^{-1}_{\gamma_{1}\gamma_{2}\gamma_{3};\epsilon_{1}\epsilon_{2}\epsilon_{3}}(k_{1},k_{2},k_{3})\  F^{(3)}_{\epsilon_{1}\epsilon_{2}\epsilon_{3};\beta_{1}\beta_{2}\beta_{3}}(k_{1},k_{2},k_{3},z) = 0,
    \end{split}
\end{equation}
which, after converting back to the time domain, yields an equation of motion that is similar in structure to the one derived for the 4-point density correlation function:
\begin{equation}\label{eomF3}
    \begin{split}
    &\pd{}{t}F^{(3)}_{\alpha_{1}\alpha_{2}\alpha_{3};\beta_{1}\beta_{2}\beta_{3}}(k_{1},k_{2},k_{3},t) + H^{(3)}_{\alpha_{1}\alpha_{2}\alpha_{3};\gamma_{1}\gamma_{2}\gamma_{3}}(k_{1},k_{2},k_{3}) \left(S^{(3)}\right)^{-1}_{\gamma_{1}\gamma_{2}\gamma_{3};\epsilon_{1}\epsilon_{2}\epsilon_{3}}(k_{1},k_{2},k_{3})\  F^{(3)}_{\epsilon_{1}\epsilon_{2}\epsilon_{3};\beta_{1}\beta_{2}\beta_{3}}(k_{1},k_{2},k_{3},t)\\
    &\quad\quad + \int_{0}^{t} dt^{\prime}  M^{(3)}_{\alpha_{1}\alpha_{2}\alpha_{3};\gamma_{1}\gamma_{2}\gamma_{3}}(k_{1},k_{2},k_{3},t-t^{\prime}) \left(H^{(3)}\right)^{-1}_{\gamma_{1}\gamma_{2}\gamma_{3};\epsilon_{1}\epsilon_{2}\epsilon_{3}}(k_{1},k_{2},k_{3})\  \pd{}{t^{\prime}}F^{(3)}_{\epsilon_{1}\epsilon_{2}\epsilon_{3};\beta_{1}\beta_{2}\beta_{3}}(k_{1},k_{2},k_{3},t^{\prime}) =0. 
    \end{split}
\end{equation}

\subsection*{Simplification of the six-point memory kernel}
\noindent In the main text we have explicitly demonstrated how the irreducible 4-point memory kernel, \cref{M2_formal}, can be simplified to \cref{M2final} by projecting the fluctuating forces on density triplets. Here we show that a similar procedure can be carried out for the irreducible 6-point density correlation function. Converting \cref{M3_formal} back to the time domain we have as its definition
\begin{equation}\label{M3_formal2}
    M^{(3)}_{\alpha_{1}\alpha_{2}\alpha_{3};\beta_{1}\beta_{2}\beta_{3}}(k_{1},k_{2},k_{3},z)=\avg{\rho_{\alpha_{1}}^{*} (\mv{k}_{1})\rho_{\alpha_{2}}^{*} (\mv{k}_{2})\rho_{\alpha_{3}}^{*} (\mv{k}_{3}) \Omega \mathcal{Q}^{(3)} e^{\Omega^{(3)}_{\mathrm{irr}}t} \mathcal{Q}^{(3)} \Omega \rho_{\beta_{1}} (\mv{k}_{1})\rho_{\beta_{2}} (\mv{k}_{2})\rho_{\beta_{3}} (\mv{k}_{3})}.
\end{equation}
The irreducible $6$-point memory kernel is now represented by an irreducible correlation of third order fluctuating forces ($\mathcal{Q}^{(3)}\Omega\rho_{\beta_{1}} (\mv{k}_{1})\rho_{\beta_{2}} (\mv{k}_{2})\rho_{\beta_{3}} (\mv{k}_{3})\rangle$). Since these forces are, to leading order, dominated by products of four density modes~\cite{Janssen2015a}, we project them on the space of density quadruplets. Such a projection can be carried out in the same spirit as before, namely by placing the operator 
\begin{equation}
    \mathcal{P}_{4}=\frac{1}{24}\sum_{\mv{q}_{1},\mv{q}_{2},\mv{q}_{3},\mv{q}_{4}} | \rho_{\mu_{1}} (\mv{q}_{1})\rho_{\mu_{2}} (\mv{q}_{2}) \rho_{\mu_{3}} (\mv{q}_{3})\rho_{\mu_{4}} (\mv{q}_{4}) \rangle \chi_{1234}
\langle \rho_{\nu_{1}}^{*} (\mv{q}_{1})\rho_{\nu_{2}}^{*} (\mv{q}_{2}) \rho_{\nu_{3}}^{*} (\mv{q}_{3}) \rho_{\nu_{4}}^{*} (\mv{q}_{4})|,
\end{equation}
in front of both forces. Note that this projector is  normalized by $\chi_{1234}=S^{-1}_{\mu_{1};\nu_{1}}(q_{1})  S^{-1}_{\mu_{2};\nu_{2}}(q_{2})
S^{-1}_{\mu_{3};\nu_{3}}(q_{3})S^{-1}_{\mu_{4};\nu_{4}}(q_{4})$ and has a prefactor to compensate for double counting. As a result, the memory kernel is again rewritten in terms of static 'vertices' encompassing a higher order dynamic correlation function: 
\begin{equation}
\begin{split}
    M^{(3)}_{\alpha_{1}\alpha_{2}\alpha_{3};\beta_{1}\beta_{2}\beta_{3}} (k_{1},k_{2},k_{3},t) \approx \frac{1}{576} & \sum_{\mv{q}_{1}...\mv{q}_{8}} \avg{\rho_{\alpha_{1}}^{*} (\mv{k}_{1})\rho_{\alpha_{2}}^{*} (\mv{k}_{2})\rho_{\alpha_{3}}^{*} (\mv{k}_{3})\Omega \mathcal{Q}^{(3)} \rho_{\mu_{1}} (\mv{q}_{1})\rho_{\mu_{2}} (\mv{q}_{2}) \rho_{\mu_{3}}(\mv{q}_{3})\rho_{\mu_{4}}(\mv{q}_{4})}\chi_{1234} \\ & \quad \quad \ \ \avg{\rho_{\nu_{1}}^{*} (\mv{q}_{1})\rho_{\nu_{2}}^{*} (\mv{q}_{2}) \rho_{\nu_{3}}^{*} (\mv{q}_{3})\rho_{\nu_{4}}^{*} (\mv{q}_{4})e^{\Omega^{(3)}_{\mathrm{irr}}t} \rho_{\mu_{5}} (\mv{q}_{5})\rho_{\mu_{6}} (\mv{q}_{6}) \rho_{\mu_{7}} (\mv{q}_{7})\rho_{\mu_{8}} (\mv{q}_{8})}
     \\ 
    & \quad \quad \quad \chi_{5678} \avg{\rho^{*}_{\nu_{5}} (\mv{q}_{5})\rho^{*}_{\nu_{6}} (\mv{q}_{6}) \rho^{*}_{\nu_{7}}(\mv{q}_{7})\rho^{*}_{\nu_{8}}(\mv{q}_{8})\Omega \mathcal{Q}^{(3)} \rho_{\beta_{1}} (\mv{k}_{1})\rho_{\beta_{2}} (\mv{k}_{2}) \rho_{\beta_{3}} (\mv{k}_{3})}.
\end{split}
\end{equation}
To make this function manageable we work out both vertices (see below for more details), i.e.\  
\begin{equation}\label{vertex3}
\begin{split}
    & \avg{\rho_{\alpha_{1}}^{*} (\mv{k}_{1})\rho_{\alpha_{2}}^{*} (\mv{k}_{2})\rho_{\alpha_{3}}^{*} (\mv{k}_{3})\Omega \mathcal{Q}^{(3)} \rho_{\mu_{1}} (\mv{q}_{1})\rho_{\mu_{2}} (\mv{q}_{2}) \rho_{\mu_{3}}(\mv{q}_{3})\rho_{\mu_{4}}(\mv{q}_{4})}\chi_{1234} = \Bigg[ \frac{D^{\alpha_{1}}}{\sqrt{N_{\alpha_{1}}}} \bigg(  \delta_{\mv{k}_{1},\mv{q}_{1}+\mv{q}_{2}}\delta_{\mv{k}_{2},\mv{q}_{3}}\delta_{\mv{k}_{3},\mv{q}_{4}} \delta_{\alpha_{2}\nu_{3}} \delta_{\alpha_{3}\nu_{4}}  \\
    & \qquad \qquad \qquad \qquad \qquad \Big(  k^{2}_{1} \delta_{\alpha_{1}\nu_{1}} \delta_{\alpha_{1}\nu_{2}} + 
     \mv{k}_{1}\cdot\mv{q}_{1} \delta_{\alpha_{1}\nu_{2}}
    S^{-1}_{\alpha_{1};\nu_{1}}(q_{1})
    +\ \mv{k}_{1}\cdot\mv{q}_{2} \delta_{\alpha_{1}\nu_{1}} S^{-1}_{\alpha_{1};\nu_{2}}(q_{2}) \Big) +\ \{ \mv{q}_{3},\nu_{3} \leftrightarrow \mv{q}_{4},\nu_{4}\} \bigg)  \\
    & \qquad \qquad \qquad \qquad \qquad + \{ \mv{q}_{1},\nu_{1} \leftrightarrow \mv{q}_{3},\nu_{3}\} + \{ \mv{q}_{2},\nu_{2} \leftrightarrow \mv{q}_{3},\nu_{3}\} + \{ \mv{q}_{1},\nu_{1} \leftrightarrow \mv{q}_{4},\nu_{4}\} + \{ \mv{q}_{2},\nu_{2} \leftrightarrow \mv{q}_{4},\nu_{4}\} \\
    & \qquad \qquad \qquad \qquad \qquad + \{ \mv{q}_{1,2},\nu_{1,2} \leftrightarrow \mv{q}_{3,4},\nu_{3,4}\} \Bigg] + \{ \mv{k}_{1},\alpha_{1} \leftrightarrow \mv{k}_{2},\alpha_{2}\} + \{ \mv{k}_{1},\alpha_{1} \leftrightarrow \mv{k}_{3},\alpha_{3}\},
\end{split}
\end{equation}
and likewise for the right vertex. Comparing \cref{vertex3,vertex2} we find that the vertices of the third order memory kernel remain almost the same as those of the second order one, except for the addition of several symmetric terms due to additional wavevector couplings. To illustrate what this implies for the memory kernel, we utilize the vertices, work out the Kronecker deltas, and collect symmetric terms, so that it can be written as
\begin{equation}
\begin{split}
     M^{(3)}_{\alpha_{1}\alpha_{2}\alpha_{3};\beta_{1}\beta_{2}\beta_{3}} & (k_{1},k_{2},k_{3},t)  \approx  \frac{1}{4} \sum_{\mv{q},\mv{q}^{\prime}} \Bigg[ \frac{D^{\alpha_{1}}}{\sqrt{N_{\alpha_{1}}}} V^{\alpha_{1}}_{\mu\nu}(\mv{k}_{1},\mv{q}) \\
    &  \bigg( \avg{ \rho_{\mu}^{*} (\mv{q})\rho_{\nu}^{*} (\mv{k}_{1}-\mv{q})\rho_{\alpha_{2}}^{*} (\mv{k}_{2})\rho_{\alpha_{3}}^{*} (\mv{k}_{3}) e^{\Omega^{(3)}_{\mathrm{irr}} t} \rho_{\mu^{\prime}} (\mv{q}^{\prime})\rho_{\nu^{\prime}} (\mv{k}_{1}-\mv{q}^{\prime})\rho_{\beta_{2}} (\mv{k}_{2})\rho_{\beta_{3}} (\mv{k}_{3})} 
    \frac{D^{\beta_{1}}}{\sqrt{N_{\beta_{1}}}} V^{\beta_{1}}_{\mu^{\prime}\nu^{\prime}}(\mv{k}_{1},\mv{q}^{\prime})\ + \\
    & \quad \avg{\rho_{\mu}^{*} (\mv{q})\rho_{\nu}^{*} (\mv{k}_{1}-\mv{q})\rho_{\alpha_{2}}^{*} (\mv{k}_{2})\rho_{\alpha_{3}}^{*} (\mv{k}_{3}) e^{\Omega^{(3)}_{\mathrm{irr}} t} \rho_{\mu^{\prime}} (\mv{q}^{\prime})\rho_{\nu^{\prime}} (\mv{k}_{2}-\mv{q}^{\prime})\rho_{\beta_{1}} (\mv{k}_{1})\rho_{\beta_{3}} (\mv{k}_{3})} \frac{D^{\beta_{2}}}{\sqrt{N_{\beta_{2}}}} V^{\beta_{2}}_{\mu^{\prime}\nu^{\prime}}(\mv{k}_{2},\mv{q}^{\prime})\ + \\
    & \quad \avg{\rho_{\mu}^{*} (\mv{q})\rho_{\nu}^{*} (\mv{k}_{1}-\mv{q})\rho_{\alpha_{2}}^{*} (\mv{k}_{2})\rho_{\alpha_{3}}^{*} (\mv{k}_{3}) e^{\Omega^{(3)}_{\mathrm{irr}} t} \rho_{\mu^{\prime}} (\mv{q}^{\prime})\rho_{\nu^{\prime}} (\mv{k}_{3}-\mv{q}^{\prime})\rho_{\beta_{1}} (\mv{k}_{1})\rho_{\beta_{2}} (\mv{k}_{2})} \frac{D^{\beta_{3}}}{\sqrt{N_{\beta_{3}}}} V^{\beta_{3}}_{\mu^{\prime}\nu^{\prime}}(\mv{k}_{3},\mv{q}^{\prime}) \bigg) \Bigg] \\
    & \qquad + \{ \mv{k}_{1},\alpha_{1},\beta_{1} \leftrightarrow \mv{k}_{2},\alpha_{2},\beta_{2}\} + \{ \mv{k}_{1},\alpha_{1},\beta_{1} \leftrightarrow \mv{k}_{3},\alpha_{3},\beta_{3}\}. 
\end{split}
\end{equation}
After replacing the irreducible evolution operator with a full one and applying the diagonal approximation ($\mv{q}=\mv{q}^{\prime},\ \mv{k}_{1,2,3}-\mv{q}=\mv{q}^{\prime}$, retaining only diagonal correlations), the memory kernel is simplified to 
\begin{equation}\label{M3final}
\begin{split}
    M^{(3)}_{\alpha_{1}\alpha_{2}\alpha_{3};\beta_{1}\beta_{2}\beta_{3}} (k_{1},k_{2},k_{3},t) \approx \frac{1}{2} & \sum_{\mv{q}}  \frac{D^{\alpha_{1}}}{\sqrt{N_{\alpha_{1}}}} V^{\alpha_{1}}_{\mu\nu}(\mv{k}_{1},\mv{q}) F^{(4)}_{\mu\nu\alpha_{2}\alpha_{3};\mu^{\prime}\nu^{\prime}\beta_{2}\beta_{3}}(q,\abs{\mv{k}_{1}-\mv{q}},k_{2},k_{3},t) \frac{D^{\beta_{1}}}{\sqrt{N_{\beta_{1}}}} V^{\beta_{1}}_{\mu^{\prime}\nu^{\prime}}(\mv{k}_{1},\mv{q}) \\[3pt]
    & + \{ \mv{k}_{1},\alpha_{1},\beta_{1} \leftrightarrow \mv{k}_{2},\alpha_{2},\beta_{2}\} + \{ \mv{k}_{1},\alpha_{1},\beta_{1} \leftrightarrow \mv{k}_{3},\alpha_{3},\beta_{3}\},
\end{split}
\end{equation}
where
\begin{equation}
F^{(4)}_{\alpha_{1}\alpha_{2}\alpha_{3}\alpha_{4};\beta_{1}\beta_{2}\beta_{3}\beta_{4}}(k_{1},k_{2},k_{3},k_{4},t) =\avg{\rho_{\alpha_{1}}^{*} (\mv{k}_{1})\rho_{\alpha_{2}}^{*} (\mv{k}_{2}) \rho_{\alpha_{3}}^{*} (\mv{k}_{3})\rho_{\alpha_{4}}^{*} (\mv{k}_{4}) e^{\Omega t} \rho_{\beta_{1}} (\mv{k}_{1})\rho_{\beta_{2}} (\mv{k}_{2})\rho_{\beta_{3}} (\mv{k}_{3})\rho_{\beta_{4}} (\mv{k}_{4})}
\end{equation}
is the $8$-point density correlation, which is the next correlation in the hierarchy.

\subsection*{Vertices}\label{vertex_der}
\noindent As discussed above, in order to find an approximate expression for the second and third order memory kernels we have projected the associated fluctuating forces onto density triplets and quadruplets, respectively. Consequently, the memory kernels are represented by two static vertices enclosing a higher order correlation function. For completeness, we now present a more detailed derivation of the left vertices (the right vertices can be found in an identical manner) for both mentioned memory kernels. 

\subsubsection*{Four-point memory kernel}
\noindent As our starting point we take the definition of the left vertex presented in the main text, i.e.\
\begin{equation}\label{leftvertexM2}
\begin{split}
    & \hspace{-1.5cm} \avg{\rho_{\alpha_{1}}^{*} (\mv{k}_{1})\rho_{\alpha_{2}}^{*} (\mv{k}_{2}) \Omega \mathcal{Q}^{(2)} \rho_{\mu_{1}} (\mv{q}_{1})\rho_{\mu_{2}} (\mv{q}_{2}) \rho_{\mu_{3}}(\mv{q}_{3})} S^{-1}_{\mu_{1};\nu_{1}}(q_{1})  S^{-1}_{\mu_{2};\nu_{2}}(q_{2})
S^{-1}_{\mu_{3};\nu_{3}}(q_{3}) \\[4pt]
& \hspace{0.0cm} = \avg{\rho_{\alpha_{1}}^{*} (\mv{k}_{1})\rho_{\alpha_{2}}^{*} (\mv{k}_{2}) \Omega \rho_{\mu_{1}} (\mv{q}_{1})\rho_{\mu_{2}} (\mv{q}_{2}) \rho_{\mu_{3}}(\mv{q}_{3})} S^{-1}_{\mu_{1};\nu_{1}}(q_{1})  S^{-1}_{\mu_{2};\nu_{2}}(q_{2})
S^{-1}_{\mu_{3};\nu_{3}}(q_{3}) \\[4pt]  
& \hspace{0.5cm} - \avg{\rho_{\alpha_{1}}^{*} (\mv{k}_{1})\rho_{\alpha_{2}}^{*} (\mv{k}_{2}) \Omega \mathcal{P}^{(2)} \rho_{\mu_{1}} (\mv{q}_{1})\rho_{\mu_{2}} (\mv{q}_{2}) \rho_{\mu_{3}}(\mv{q}_{3})} S^{-1}_{\mu_{1};\nu_{1}}(q_{1})  S^{-1}_{\mu_{2};\nu_{2}}(q_{2})
S^{-1}_{\mu_{3};\nu_{3}}(q_{3}),
\end{split}
\end{equation}
and first focus on the second term, which can be written as
\begin{equation}\label{leftvertexM2_2}
    H^{(2)}_{\alpha_{1}\alpha_{2};\gamma_{1}\gamma_{2}}(k_{1},k_{2}) S^{-1}_{\gamma_{1};\epsilon_{1}}(k_{1})  S^{-1}_{\gamma_{2};\epsilon_{2}}(k_{2}) \avg{\rho_{\epsilon_{1}}^{*} (\mv{k}_{1})\rho_{\epsilon_{2}}^{*} (\mv{k}_{2}) \rho_{\mu_{1}} (\mv{q}_{1})\rho_{\mu_{2}} (\mv{q}_{2}) \rho_{\mu_{3}}(\mv{q}_{3})}  S^{-1}_{\mu_{1};\nu_{1}}(q_{1})  S^{-1}_{\mu_{2};\nu_{2}}(q_{2}) S^{-1}_{\mu_{3};\nu_{3}}(q_{3}).
\end{equation}
The $5$-point static structure correlation can then be approximated using Gaussian factorization and the convolution approximation, yielding 
\begin{equation}
\begin{split}
    & \hspace{-1.2cm} \avg{\rho_{\epsilon_{1}}^{*} (\mv{k}_{1})\rho_{\epsilon_{2}}^{*} (\mv{k}_{2}) \rho_{\mu_{1}} (\mv{q}_{1})\rho_{\mu_{2}} (\mv{q}_{2}) \rho_{\mu_{3}}(\mv{q}_{3})} \approx \bigg( \delta_{\mv{k}_{1},\mv{q}_{1}+\mv{q}_{2}}\delta_{\mv{k}_{2},\mv{q}_{3}} \frac{1}{\sqrt{N_{\lambda}}} S_{\epsilon_{1};\lambda}(k_{1}) S_{\lambda;\mu_{1}}(q_{1}) S_{\lambda;\mu_{2}}(q_{2}) S_{\epsilon_{2};\mu_{3}}(q_{3}) \\[4pt]
    & \hspace{4.9cm} + \{ \mv{q}_{1},\mu_{1} \leftrightarrow \mv{q}_{3},\mu_{3}\} + \{ \mv{q}_{2},\mu_{2} \leftrightarrow \mv{q}_{3},\mu_{3}\} \bigg)  + \{ \mv{k}_{1},\epsilon_{1} \leftrightarrow \mv{k}_{2},\epsilon_{2}\}.
\end{split}
\end{equation}
Inserting this expression into \cref{leftvertexM2_2} allows us to simplify the second term as 
\begin{equation}\label{leftvertexM2_2s}
\begin{split}
    & \hspace{-2.0cm} \bigg[ \delta_{\mv{k}_{1},\mv{q}_{1}+\mv{q}_{2}}\delta_{\mv{k}_{2},\mv{q}_{3}} \frac{1}{\sqrt{N_{\lambda}}} \Big( k_{1}^{2} D^{\alpha_{1}}\delta_{\alpha_{1}\lambda}\delta_{\alpha_{2}\nu_{3}} + k_{2}^{2} D^{\alpha_{2}} S_{\alpha_{1};\lambda}(k_{1}) S^{-1}_{\alpha_{2};\nu_{3}}(k_{2}) \Big) \delta_{\lambda\nu_{1}}\delta_{\lambda\nu_{2}} \\[4pt]
    & \hspace{0.0cm} + \{ \mv{q}_{1},\nu_{1} \leftrightarrow \mv{q}_{3},\nu_{3}\} + \{ \mv{q}_{2},\nu_{2} \leftrightarrow \mv{q}_{3},\nu_{3}\} \bigg]  + \{ \mv{k}_{1},\alpha_{1} \leftrightarrow \mv{k}_{2},\alpha_{2}\}.
\end{split}
\end{equation}
Proceeding to the first term in \cref{leftvertexM2}, we may evaluate the evolution operator $\Omega$, which results in 
\begin{equation}
\begin{split}
    & \hspace{-1.0cm} - \bigg[ \Big(\frac{D^{\alpha_{1}}}{\sqrt{N_{\alpha_{1}}}}  \delta_{\alpha_{1}\mu_{1}} \mv{k}_{1} \cdot \mv{q}_{1} \avg{\rho_{\alpha_{1}}^{*} (\mv{k}_{1}-\mv{q}_{1})\rho_{\alpha_{2}}^{*} (\mv{k}_{2}) \rho_{\mu_{2}} (\mv{q}_{2}) \rho_{\mu_{3}}(\mv{q}_{3})} + \{ \mv{q}_{1},\mu_{1} \leftrightarrow \mv{q}_{2},\mu_{2}\} + \{ \mv{q}_{1},\mu_{1} \leftrightarrow \mv{q}_{3},\mu_{3}\} \Big) \\[4pt]
    & \hspace{2.5cm}  + \{ \mv{k}_{1},\alpha_{1} \leftrightarrow \mv{k}_{2},\alpha_{2}\} \bigg] S^{-1}_{\mu_{1};\nu_{1}}(q_{1})  S^{-1}_{\mu_{2};\nu_{2}}(q_{2})
S^{-1}_{\mu_{3};\nu_{3}}(q_{3}).
\end{split}
\end{equation}
Now we realize that (again using Gaussian factorization and the convolution approximation),  
\begin{equation}
\begin{split}
    & \hspace{-1.0cm} \avg{\rho_{\alpha_{1}}^{*} (\mv{k}_{1}-\mv{q}_{1})\rho_{\alpha_{2}}^{*} (\mv{k}_{2}) \rho_{\mu_{2}} (\mv{q}_{2}) \rho_{\mu_{3}}(\mv{q}_{3})} \approx \delta_{\mv{k}_{1},\mv{q}_{1}+\mv{q}_{2}}\delta_{\mv{k}_{2},\mv{q}_{3}}S_{\alpha_{1};\mu_{2}}(q_{2})S_{\alpha_{2};\mu_{3}}(q_{3}) \\[4pt]
    & \hspace{0.0cm} + \delta_{\mv{k}_{1},\mv{q}_{1}+\mv{q}_{3}}\delta_{\mv{k}_{2},\mv{q}_{2}}S_{\alpha_{1};\mu_{3}}(q_{3})S_{\alpha_{2};\mu_{2}}(q_{2}) + \sqrt{N_{\alpha_{1}}} \delta_{\mv{k}_{1},\mv{q}_{1}}\delta_{\mv{k}_{2},\mv{q}_{2}+\mv{q}_{3}} \frac{1}{\sqrt{N_{\lambda}}} S_{\alpha_{2};\lambda}(k_{2}) S_{\lambda;\mu_{2}}(q_{2}) S_{\lambda;\mu_{3}}(q_{3}),
\end{split}
\end{equation}
and use this result to rewrite the first term of the left vertex as follows:
\begin{equation}\label{leftvertexM2_1}
\begin{split}
    & \hspace{-0.5cm} - \bigg[ \delta_{\mv{k}_{1},\mv{q}_{1}+\mv{q}_{2}}\delta_{\mv{k}_{2},\mv{q}_{3}} \Big( \frac{D^{\alpha_{1}}}{\sqrt{N_{\alpha_{1}}}} \mv{k}_{1} \cdot \mv{q}_{1} \delta_{\alpha_{1}\nu_{2}}\delta_{\alpha_{2}\nu_{3}} S^{-1}_{\alpha_{1};\nu_{1}}(q_{1}) + \frac{D^{\alpha_{1}}}{\sqrt{N_{\alpha_{1}}}} \mv{k}_{1} \cdot \mv{q}_{2} \delta_{\alpha_{1}\nu_{1}}\delta_{\alpha_{2}\nu_{3}} S^{-1}_{\alpha_{1};\nu_{2}}(q_{2}) \\[4pt]
    & \hspace{0.5cm} + \frac{D^{\alpha_{2}}}{\sqrt{N_{\lambda}}} k_{2}^{2} S_{\alpha_{1};\lambda}(k_{1}) S^{-1}_{\alpha_{2};\nu_{3}}(k_{2}) \delta_{\lambda\nu_{1}}\delta_{\lambda\nu_{2}} \Big) + \{ \mv{q}_{1},\nu_{1} \leftrightarrow \mv{q}_{3},\nu_{3}\} + \{ \mv{q}_{2},\nu_{2} \leftrightarrow \mv{q}_{3},\nu_{3}\} \bigg]  + \{ \mv{k}_{1},\alpha_{1} \leftrightarrow \mv{k}_{2},\alpha_{2}\}.
\end{split}
\end{equation}
Combining both \cref{leftvertexM2_2s,leftvertexM2_1} then results in
\begin{equation}
\begin{split}
    & \hspace{-1.0cm} \avg{\rho_{\alpha_{1}}^{*} (\mv{k}_{1})\rho_{\alpha_{2}}^{*} (\mv{k}_{2}) \Omega \mathcal{Q}^{(2)} \rho_{\mu_{1}} (\mv{q}_{1})\rho_{\mu_{2}} (\mv{q}_{2}) \rho_{\mu_{3}}(\mv{q}_{3})} S^{-1}_{\mu_{1};\nu_{1}}(q_{1})  S^{-1}_{\mu_{2};\nu_{2}}(q_{2})
S^{-1}_{\mu_{3};\nu_{3}}(q_{3}) = \\[4pt]
& \hspace{0.5cm} \Bigg[ \frac{D^{\alpha_{1}}}{\sqrt{N_{\alpha_{1}}}} \bigg( \delta_{\mv{k}_{1},\mv{q}_{1}+\mv{q}_{2}}\delta_{\mv{k}_{2},\mv{q}_{3}} \delta_{\alpha_{2}\nu_{3}}  \Big(  k^{2}_{1} \delta_{\alpha_{1}\nu_{1}} \delta_{\alpha_{1}\nu_{2}} +  \mv{k}_{1}\cdot\mv{q}_{1} \delta_{\alpha_{1}\nu_{2}}
    S^{-1}_{\alpha_{1};\nu_{1}}(q_{1})
    +\ \mv{k}_{1}\cdot\mv{q}_{2} \delta_{\alpha_{1}\nu_{1}} S^{-1}_{\alpha_{1};\nu_{2}}(q_{2}) \Big) \bigg) \\[4pt]
    & \hspace{4.0cm} \ +\ \{ \mv{q}_{1},\nu_{1} \leftrightarrow \mv{q}_{3},\nu_{3}\} + \{ \mv{q}_{2},\nu_{2} \leftrightarrow \mv{q}_{3},\nu_{3}\} \Bigg] + \{ \mv{k}_{1},\alpha_{1} \leftrightarrow \mv{k}_{2},\alpha_{2}\},
\end{split}
\end{equation}
which is the same expression as presented in the main text.

\subsubsection*{Six-point memory kernel}
\noindent For the $6$-point memory kernel, the left vertex is defined as
\begin{equation}\label{leftvertexM3}
\begin{split}
    & \hspace{-0.0cm} \avg{\rho_{\alpha_{1}}^{*} (\mv{k}_{1})\rho_{\alpha_{2}}^{*} (\mv{k}_{2}) \rho_{\alpha_{3}}^{*} (\mv{k}_{3}) \Omega \mathcal{Q}^{(3)} \rho_{\mu_{1}} (\mv{q}_{1})\rho_{\mu_{2}} (\mv{q}_{2}) \rho_{\mu_{3}}(\mv{q}_{3})\rho_{\mu_{4}}(\mv{q}_{4})} S^{-1}_{\mu_{1};\nu_{1}}(q_{1})  S^{-1}_{\mu_{2};\nu_{2}}(q_{2})
S^{-1}_{\mu_{3};\nu_{3}}(q_{3}) S^{-1}_{\mu_{4};\nu_{4}}(q_{4}) \\[4pt]
& \hspace{0.5cm} = \avg{\rho_{\alpha_{1}}^{*} (\mv{k}_{1})\rho_{\alpha_{2}}^{*} (\mv{k}_{2}) \rho_{\alpha_{3}}^{*} (\mv{k}_{3}) \Omega \rho_{\mu_{1}} (\mv{q}_{1})\rho_{\mu_{2}} (\mv{q}_{2}) \rho_{\mu_{3}}(\mv{q}_{3})\rho_{\mu_{4}}(\mv{q}_{4})} S^{-1}_{\mu_{1};\nu_{1}}(q_{1})  S^{-1}_{\mu_{2};\nu_{2}}(q_{2})
S^{-1}_{\mu_{3};\nu_{3}}(q_{3}) S^{-1}_{\mu_{4};\nu_{4}}(q_{4}) \\[4pt]  
& \hspace{0.8cm} - \avg{\rho_{\alpha_{1}}^{*} (\mv{k}_{1})\rho_{\alpha_{2}}^{*} (\mv{k}_{2}) \rho_{\alpha_{3}}^{*} (\mv{k}_{3}) \Omega \mathcal{P}^{(3)} \rho_{\mu_{1}} (\mv{q}_{1})\rho_{\mu_{2}} (\mv{q}_{2}) \rho_{\mu_{3}}(\mv{q}_{3})\rho_{\mu_{4}}(\mv{q}_{4})} S^{-1}_{\mu_{1};\nu_{1}}(q_{1})  S^{-1}_{\mu_{2};\nu_{2}}(q_{2})
S^{-1}_{\mu_{3};\nu_{3}}(q_{3}) S^{-1}_{\mu_{4};\nu_{4}}(q_{4}).
\end{split}
\end{equation}
Using the fact that the $7$-point static structure correlation and $6$-point collective diffusion tensor can be approximated (invoking Gaussian factorization and the convolution approximation) to give
\begin{equation}
\begin{split}
    & \hspace{-0.0cm} \avg{\rho_{\epsilon_{1}}^{*} (\mv{k}_{1})\rho_{\epsilon_{2}}^{*} (\mv{k}_{2}) \rho_{\epsilon_{3}}^{*} (\mv{k}_{3}) \rho_{\mu_{1}} (\mv{q}_{1})\rho_{\mu_{2}} (\mv{q}_{2}) \rho_{\mu_{3}}(\mv{q}_{3})\rho_{\mu_{4}} (\mv{q}_{4})} \approx \bigg[ \Big( \delta_{\mv{k}_{1},\mv{q}_{1}+\mv{q}_{2}}\delta_{\mv{k}_{2},\mv{q}_{3}}\delta_{\mv{k}_{3},\mv{q}_{4}} \frac{1}{\sqrt{N_{\lambda}}} S_{\epsilon_{1};\lambda}(k_{1}) S_{\lambda;\mu_{1}}(q_{1}) \\[4pt]
    & \hspace{0.0cm} S_{\lambda;\mu_{2}}(q_{2}) S_{\epsilon_{2};\mu_{3}}(q_{3}) S_{\epsilon_{3};\mu_{4}}(q_{4}) + \{ \mv{q}_{3},\mu_{3} \leftrightarrow \mv{q}_{4},\mu_{4}\} \Big) + \{ \mv{q}_{1},\mu_{1} \leftrightarrow \mv{q}_{3},\mu_{3}\} + \{ \mv{q}_{1},\mu_{1} \leftrightarrow \mv{q}_{4},\mu_{4}\} + \{ \mv{q}_{2},\mu_{2} \leftrightarrow \mv{q}_{3},\mu_{3}\} \\[4pt]
    & \hspace{0.0cm} + \{ \mv{q}_{2},\mu_{2} \leftrightarrow \mv{q}_{4},\mu_{4}\} + \{ \mv{q}_{1,2},\mu_{1,2} \leftrightarrow \mv{q}_{3,4},\mu_{3,4}\} \bigg] + \{ \mv{k}_{1},\epsilon_{1} \leftrightarrow \mv{k}_{2},\epsilon_{2}\} + \{ \mv{k}_{1},\epsilon_{1} \leftrightarrow \mv{k}_{3},\epsilon_{3}\},
\end{split}
\end{equation}
and
\begin{equation}
\begin{split}
    H^{(3)}_{\alpha_{1}\alpha_{2}\alpha_{3};\beta_{1}\beta_{2}\beta_{3}} (k_{1},k_{2},k_{3}) & = - \avg{\rho_{\alpha_{1}}^{*} (\mv{k}_{1})\rho_{\alpha_{2}}^{*} (\mv{k}_{2}) \rho_{\alpha_{1}}^{*} (\mv{k}_{3})\Omega \rho_{\beta_{1}} (\mv{k}_{1})\rho_{\beta_{2}} (\mv{k}_{2})\rho_{\beta_{3}} (\mv{k}_{3})} \\[4pt]
    & \hspace{-3.0cm} \approx k_{1}^{2}D^{\alpha_{1}}\delta_{\alpha_{1}\beta_{1}}S_{\alpha_{2};\beta_{2}}(k_{2})S_{\alpha_{3};\beta_{3}}(k_{3})+k_{2}^{2}D^{\alpha_{2}}\delta_{\alpha_{2}\beta_{2}}S_{\alpha_{1};\beta_{1}}(k_{1})S_{\alpha_{3};\beta_{3}}(k_{3})+k_{3}^{2}D_{t}^{\alpha_{3}}\delta_{\alpha_{3}\beta_{3}}S_{\alpha_{1};\beta_{1}}(k_{1})S_{\alpha_{2};\beta_{2}}(k_{2}),
\end{split}
\end{equation}
respectively, allows us to write the second term in \cref{leftvertexM3} as
\begin{equation}\label{leftvertexM3_2s}
\begin{split}
    & \hspace{-0.0cm} \Bigg[ \bigg( \delta_{\mv{k}_{1},\mv{q}_{1}+\mv{q}_{2}}\delta_{\mv{k}_{2},\mv{q}_{3}}\delta_{\mv{k}_{3},\mv{q}_{4}} \frac{1}{\sqrt{N_{\lambda}}} \Big( k_{1}^{2} D^{\alpha_{1}}\delta_{\alpha_{1}\lambda}\delta_{\alpha_{2}\nu_{3}} \delta_{\alpha_{3}\nu_{4}} + k_{2}^{2} D^{\alpha_{2}} S_{\alpha_{1};\lambda}(k_{1}) S^{-1}_{\alpha_{2};\nu_{3}}(k_{2}) \delta_{\alpha_{3}\nu_{4}} + k_{3}^{2} D_{t}^{\alpha_{3}} S_{\alpha_{1};\lambda}(k_{1}) S^{-1}_{\alpha_{3};\nu_{4}}(k_{3}) \delta_{\alpha_{2}\nu_{3}} \Big) \\[4pt]
    & \hspace{0.0cm}  \delta_{\lambda\nu_{1}}\delta_{\lambda\nu_{2}} + \{ \mv{q}_{3},\nu_{3} \leftrightarrow \mv{q}_{4},\nu_{4}\} \bigg) + \{ \mv{q}_{1},\nu_{1} \leftrightarrow \mv{q}_{3},\nu_{3}\} + \{ \mv{q}_{1},\nu_{1} \leftrightarrow \mv{q}_{4},\nu_{4}\} + \{ \mv{q}_{2},\nu_{2} \leftrightarrow \mv{q}_{3},\nu_{3}\} + \{ \mv{q}_{2},\nu_{2} \leftrightarrow \mv{q}_{4},\nu_{4}\} \\[4pt]
    & \hspace{0.0cm}  + \{ \mv{q}_{1,2},\mu_{1,2} \leftrightarrow \mv{q}_{3,4},\mu_{3,4}\} \Bigg]  + \{ \mv{k}_{1},\alpha_{1} \leftrightarrow \mv{k}_{2},\alpha_{2}\} + \{ \mv{k}_{1},\alpha_{1} \leftrightarrow \mv{k}_{3},\alpha_{3}\}.
\end{split}
\end{equation}
Next, we progress to the first term in \cref{leftvertexM3} which, after evaluating the evolution operator $\Omega$, yields
\begin{equation}
\begin{split}
    & \hspace{-0.0cm} - \bigg[ \Big(\frac{D^{\alpha_{1}}}{\sqrt{N_{\alpha_{1}}}}  \delta_{\alpha_{1}\mu_{1}} \mv{k}_{1} \cdot \mv{q}_{1} \avg{\rho_{\alpha_{1}}^{*} (\mv{k}_{1}-\mv{q}_{1})\rho_{\alpha_{2}}^{*} (\mv{k}_{2}) \rho_{\alpha_{3}}^{*} (\mv{k}_{3}) \rho_{\mu_{2}} (\mv{q}_{2}) \rho_{\mu_{3}}(\mv{q}_{3}) \rho_{\mu_{4}}(\mv{q}_{4})} + \{ \mv{q}_{1},\mu_{1} \leftrightarrow \mv{q}_{2},\mu_{2}\} + \{ \mv{q}_{1},\mu_{1} \leftrightarrow \mv{q}_{3},\mu_{3}\}  \\[4pt]
    & \hspace{0.5cm} + \{ \mv{q}_{1},\mu_{1} \leftrightarrow \mv{q}_{4},\mu_{4}\} \Big) + \{ \mv{k}_{1},\alpha_{1} \leftrightarrow \mv{k}_{2},\alpha_{2}\} + \{ \mv{k}_{1},\alpha_{1} \leftrightarrow \mv{k}_{3},\alpha_{3}\} \bigg] S^{-1}_{\mu_{1};\nu_{1}}(q_{1})  S^{-1}_{\mu_{2};\nu_{2}}(q_{2})
S^{-1}_{\mu_{3};\nu_{3}}(q_{3})S^{-1}_{\mu_{4};\nu_{4}}(q_{4}).
\end{split}
\end{equation}
Applying Gaussian factorization and the convolution approximation, we have for the involved correlation function
\begin{equation}
\begin{split}
    & \hspace{-0.0cm} \avg{\rho_{\alpha_{1}}^{*} (\mv{k}_{1}-\mv{q}_{1})\rho_{\alpha_{2}}^{*} (\mv{k}_{2}) \rho_{\alpha_{3}}^{*} (\mv{k}_{3}) \rho_{\mu_{2}} (\mv{q}_{2}) \rho_{\mu_{3}}(\mv{q}_{3}) \rho_{\mu_{4}}(\mv{q}_{4})} \approx \bigg[ \Big( \delta_{\mv{k}_{1},\mv{q}_{1}+\mv{q}_{2}}\delta_{\mv{k}_{2},\mv{q}_{3}}\delta_{\mv{k}_{3},\mv{q}_{4}} S_{\alpha_{1};\mu_{2}}(q_{2})S_{\alpha_{2};\mu_{3}}(q_{3})S_{\alpha_{3};\mu_{4}}(q_{4}) \\[4pt]
    & \hspace{0.0cm} +\ \{ \mv{q}_{3},\mu_{3} \leftrightarrow \mv{q}_{4},\mu_{4}\} \Big) +\ \{ \mv{q}_{2},\mu_{2} \leftrightarrow \mv{q}_{3},\mu_{3}\} +\ \{ \mv{q}_{2},\mu_{2} \leftrightarrow \mv{q}_{4},\mu_{4}\} \bigg] + \bigg[ \Big( \sqrt{N_{\alpha_{1}}} \delta_{\mv{k}_{1},\mv{q}_{1}}\delta_{\mv{k}_{2},\mv{q}_{2}+\mv{q}_{3}}\delta_{\mv{k}_{3},\mv{q}_{4}} \\[4pt]
    &  \frac{1}{\sqrt{N_{\lambda}}} S_{\alpha_{2};\lambda}(k_{2}) S_{\lambda;\mu_{2}}(q_{2}) S_{\lambda;\mu_{3}}(q_{3}) S_{\alpha_{3};\mu_{4}}(q_{4}) +\ \{ \mv{q}_{2},\mu_{2} \leftrightarrow \mv{q}_{3},\mu_{3}\} +\ \{ \mv{q}_{2},\mu_{2} \leftrightarrow \mv{q}_{4},\mu_{4}\} \Big) + \{ \mv{k}_{2},\alpha_{2} \leftrightarrow \mv{k}_{3},\alpha_{3}\} \bigg],
\end{split}
\end{equation}
which in turn can be used to rewrite the first term of \cref{leftvertexM3} as 
\begin{equation}\label{leftvertexM3_1}
\begin{split}
    & \hspace{-0.5cm} - \Bigg[ \bigg( \delta_{\mv{k}_{1},\mv{q}_{1}+\mv{q}_{2}}\delta_{\mv{k}_{2},\mv{q}_{3}} \delta_{\mv{k}_{3},\mv{q}_{4}} \Big( \frac{D^{\alpha_{1}}}{\sqrt{N_{\alpha_{1}}}} \mv{k}_{1} \cdot \mv{q}_{1} \delta_{\alpha_{1}\nu_{2}}\delta_{\alpha_{2}\nu_{3}} \delta_{\alpha_{3}\nu_{4}} S^{-1}_{\alpha_{1};\nu_{1}}(q_{1}) + \frac{D^{\alpha_{1}}}{\sqrt{N_{\alpha_{1}}}} \mv{k}_{1} \cdot \mv{q}_{2} \delta_{\alpha_{1}\nu_{1}}\delta_{\alpha_{2}\nu_{3}} \delta_{\alpha_{3}\nu_{4}} S^{-1}_{\alpha_{1};\nu_{2}}(q_{2}) \\[4pt]
    & \hspace{0.0cm} + \frac{D^{\alpha_{2}}}{\sqrt{N_{\lambda}}} k_{2}^{2} S_{\alpha_{1};\lambda}(k_{1}) S^{-1}_{\alpha_{2};\nu_{3}}(k_{2}) \delta_{\lambda\nu_{1}}\delta_{\lambda\nu_{2}} \delta_{\alpha_{3}\nu_{4}} + \frac{D^{\alpha_{2}}}{\sqrt{N_{\lambda}}} k_{3}^{2} S_{\alpha_{1};\lambda}(k_{1}) S^{-1}_{\alpha_{3};\nu_{4}}(k_{3}) \delta_{\lambda\nu_{1}}\delta_{\lambda\nu_{2}} \delta_{\alpha_{2}\nu_{3}} \Big) + \{ \mv{q}_{3},\nu_{3} \leftrightarrow \mv{q}_{4},\nu_{4}\} \bigg) \\[4pt]
    & + \{ \mv{q}_{1},\nu_{1} \leftrightarrow \mv{q}_{3},\nu_{3}\} + \{ \mv{q}_{1},\nu_{1} \leftrightarrow \mv{q}_{4},\nu_{4}\} + \{ \mv{q}_{2},\nu_{2} \leftrightarrow \mv{q}_{3},\nu_{3}\} + \{ \mv{q}_{2},\nu_{2} \leftrightarrow \mv{q}_{4},\nu_{4}\} + \{ \mv{q}_{1,2},\nu_{1,2} \leftrightarrow \mv{q}_{3,4},\nu_{3,4}\} \Bigg] \\[4pt]
    & + \{ \mv{k}_{1},\alpha_{1} \leftrightarrow \mv{k}_{2},\alpha_{2}\} + \{ \mv{k}_{1},\alpha_{1} \leftrightarrow \mv{k}_{3},\alpha_{3}\}.
\end{split}
\end{equation}
Finally, we incorporate both \cref{leftvertexM3_2s,leftvertexM3_1} into our expression for the left vertex (\cref{leftvertexM3}). This gives
\begin{equation}
\begin{split}
    & \avg{\rho_{\alpha_{1}}^{*} (\mv{k}_{1})\rho_{\alpha_{2}}^{*} (\mv{k}_{2})\rho_{\alpha_{3}}^{*} (\mv{k}_{3})\Omega \mathcal{Q}^{(3)} \rho_{\mu_{1}} (\mv{q}_{1})\rho_{\mu_{2}} (\mv{q}_{2}) \rho_{\mu_{3}}(\mv{q}_{3})\rho_{\mu_{4}}(\mv{q}_{4})} S^{-1}_{\mu_{1};\nu_{1}}(q_{1})  S^{-1}_{\mu_{2};\nu_{2}}(q_{2})
S^{-1}_{\mu_{3};\nu_{3}}(q_{3}) S^{-1}_{\mu_{4};\nu_{4}}(q_{4}) = \\[4pt]
& \Bigg[ \frac{D^{\alpha_{1}}}{\sqrt{N_{\alpha_{1}}}} \bigg(  \delta_{\mv{k}_{1},\mv{q}_{1}+\mv{q}_{2}}\delta_{\mv{k}_{2},\mv{q}_{3}}\delta_{\mv{k}_{3},\mv{q}_{4}} \delta_{\alpha_{2}\nu_{3}} \delta_{\alpha_{3}\nu_{4}}
 \Big(  k^{2}_{1} \delta_{\alpha_{1}\nu_{1}} \delta_{\alpha_{1}\nu_{2}} + 
     \mv{k}_{1}\cdot\mv{q}_{1} \delta_{\alpha_{1}\nu_{2}}
    S^{-1}_{\alpha_{1};\nu_{1}}(q_{1})
    +\ \mv{k}_{1}\cdot\mv{q}_{2} \delta_{\alpha_{1}\nu_{1}} S^{-1}_{\alpha_{1};\nu_{2}}(q_{2}) \Big)  \\[4pt]
    & +\ \{ \mv{q}_{3},\nu_{3} \leftrightarrow \mv{q}_{4},\nu_{4}\} \bigg) + \{ \mv{q}_{1},\nu_{1} \leftrightarrow \mv{q}_{3},\nu_{3}\} + \{ \mv{q}_{2},\nu_{2} \leftrightarrow \mv{q}_{3},\nu_{3}\} + \{ \mv{q}_{1},\nu_{1} \leftrightarrow \mv{q}_{4},\nu_{4}\} + \{ \mv{q}_{2},\nu_{2} \leftrightarrow \mv{q}_{4},\nu_{4}\} \\[4pt]
    & + \{ \mv{q}_{1,2},\nu_{1,2} \leftrightarrow \mv{q}_{3,4},\nu_{3,4}\} \Bigg] + \{ \mv{k}_{1},\alpha_{1} \leftrightarrow \mv{k}_{2},\alpha_{2}\} + \{ \mv{k}_{1},\alpha_{1} \leftrightarrow \mv{k}_{3},\alpha_{3}\},
\end{split}
\end{equation}
and coincides with the result shown in the previous section of the supplementary information.

%\listoffigures

%\newpage

%\listoftables

% Figures and Tables coding should be placed where the
% first reference in the text.
% All the Figure files should be placed same working directory,
% for example (fig_1.eps and fig_1.pdf files must be present
% in the document directory)

% closing statement, nothing below matters

\end{document}